\documentclass[prd,aps,preprintnumbers,floats,floatfix,superscriptaddress,preprintnumbers,
showpacs,eqsecnum,nofootinbib,twocolumn]{revtex4-1}
\usepackage[utf8]{inputenc}
\usepackage{latexsym,array,theorem,mathrsfs,bm,float}
\usepackage{psfrag}
\usepackage{amsfonts,amsmath,amssymb,latexsym,array,afterpage,
theorem,mathrsfs,bm,float,epsfig,color,graphicx,tabularx,here,multirow}

\newcommand{\nn}{\nonumber \\}
\newcommand{\bea}{\begin{eqnarray}}
\newcommand{\ena}{\end{eqnarray}}
\newcommand{\beann}{\begin{eqnarray*}}
\newcommand{\enann}{\end{eqnarray*}}

\newcommand{\gsim}{\, \mbox{\raisebox{-1.ex}
{$\stackrel{\textstyle>}{\textstyle\sim}$}}\,}
\newcommand{\lsim}{\, \mbox{\raisebox{-1.ex}
{$\stackrel{\textstyle<}{\textstyle\sim}$}}\,}
\newcommand{\ma}[1]{\mbox{$\mathcal{#1}$}}

\newcommand{\vect}[1]{\!\!\!\mbox{\,\,\,\boldmath $#1$}}
\newcommand{\calhR}[1]{\raisebox{2ex}{\tiny ({\em h})}\hspace{-0.8em}{\ma R}}

\newcommand{\mpl}{M_{\mathrm{PL}}}

\begin{document}

\title{
Cuscuta-Galileon cosmology: \\Dynamics, gavitational ``constant''s and the Hubble constant
}


\author{Kei-ichi {\sc Maeda}}
\email{maeda-at-waseda.jp}
\address{Department of Pure and Applied Physics, Graduate School of Advanced Science and Engineering, Waseda University, 
Okubo 3-4-1, Shinjuku, Tokyo 169-8555, Japan}
\author{Sirachak {\sc Panpanich}}
\email{sirachakp-at-aoni.waseda.jp}
\address{Department of Pure and Applied Physics, Graduate School of Advanced Science and Engineering, Waseda University, 
Okubo 3-4-1, Shinjuku, Tokyo 169-8555, Japan}


\date{\today}

\begin{abstract}

We discuss cosmology based on a Cuscuta-Galileon gravity theory, 
which preserves just two degrees of freedom.
Although there exist no additional degrees of freedom, 
introduction of a potential of a scalar field changes the dynamics.
The scalar field is completely determined by matter fields.
Giving an exponential potential as an example, we discuss 
the cosmological dynamics. The gravitational ``constant'' $G_{\rm F}$ appeared
 in the effective Friedmann equation becomes time dependent.
 We also present how to construct a potential when we know the
 evolution of the  Hubble parameter. When we assume the $\Lambda$CDM 
 cosmology for the background evolution, we find the potential form.

We then analyze the density perturbations, which equation is  characterized 
only by a change of the gravitational ``constant'' $G_{\rm eff}$, which also 
becomes time dependent. 
From  the observational constraints such as 
 the constraint from the big-bang nucleosynthesis  and the constraint 
on time-variation of gravitational constant, 
we restrict the parameters in our models.

Taking into account the time dependence of the gravitational constant in the effective Friedmann equation, 
we may have a chance to explain the Hubble tension problem.

\end{abstract}


\maketitle

\section{Introduction}
\label{intro}

In order to explain the accelerated expansion of the Universe \cite{Riess:1998cb,Perlmutter:1998np} we require a mysterious energy, so-called dark energy. The dark energy candidates are a cosmological constant \cite{Weinberg:1988cp}, a scalar field \cite{Caldwell:1997ii,Copeland:1997et}, a vector field \cite{Heisenberg:2014rta,DeFelice:2016yws}, a massive tensor field \cite{deRham:2010ik,deRham:2010kj,deRham:2010tw}, or even modification on the general relativity \cite{DeFelice:2010aj}. However, until now, these candidates or deviations from general relativity (GR)  have not been detected in the solar system scale \cite{Will:2014kxa}.

To address the above problem, many modified gravity theories 
  with various types of screening mechanisms, such as chameleon screening \cite{Khoury:2003aq,Khoury:2003rn}, symmetron screening \cite{Hinterbichler:2010es}, and Vainshtein screening \cite{Vainshtein:1972sx,Nicolis:2008in}, have been proposed. These mechanisms use an effective potential to vary a mass of a scalar field \cite{Khoury:2003aq,Khoury:2003rn} or coupling between a matter field and a scalar field \cite{Hinterbichler:2010es,Brax:2010gi}, also nonlinear form of equation of motion of the scalar field leads to suppression on a fifth force \cite{Vainshtein:1972sx,Nicolis:2008in,Burrage:2014uwa,Panpanich:2019rij,Brax:2012jr,Babichev:2009ee,Babichev:2013usa}. 
  
Another solution to the previous problem is constructing new gravitational theories which propagate only two degrees of freedom as GR. 
Recently, two types of theories have been developed:
One is called minimally modified gravity \cite{Lin:2017oow,Aoki:2018zcv,Aoki:2018brq,Mukohyama:2019unx,DeFelice:2020eju,Aoki:2020oqc} whose gravitational Hamiltonian is constrained to provide only two degrees of freedom. 
The other one is called Cuscuton gravity theory \cite{Afshordi:2006ad,Afshordi:2007yx} or its extended version \cite{Iyonaga:2018vnu,Iyonaga:2020bmm}.
The extended Cuscuton theory is generalization of the original Cuscuton theory
 in the context of the beyond Horndeski theories \cite{Gleyzes:2014dya}, in which the  second-order time derivatives of a scalar field in the  equation of motion disappears, thus the scalar field is a nondynamical field. 
 Both theories have some relation as shown in the Ref. \cite{Mukohyama:2019unx}. 

In this work we consider the modified gravity with two degrees of freedom in the extended Cuscuton framework. To find cosmological solutions we have to define explicit form of theory, one example has been given in the Ref. \cite{Iyonaga:2020bmm}. We are interested in the explicit form inspired from the Cuscuta-Galileon gravity \cite{deRham:2016ged} which is a Galileon generalization of the original Cuscuton gravity. Its cosmological dynamics of the model has been studied  in Ref. \cite{Panpanich:2021lsd} where the Cuscuta-Galileon provides the sequence of the thermal history of the Universe successfully; however, the model actually has three degrees of freedom. Therefore, it is interesting to investigate cosmological solutions of the Cuscuta-Galileon gravity which has only two degrees of freedom whether the model still provides the thermal history of Universe correctly or not.

The paper is organized as follows. In \S. II, we will give our Cuscuta-Galileon gravity theory and show that it has two dynamical degrees of freedom.
In \S. III, we apply it to cosmological model and 
present the effective Friedmann equation assuming the  flat Friedmann-Lema\^{\i}tre-Robertson-Walker (FLRW) metric.
In order to study the cosmological dynamics, in \S. IV, we analyze the cosmological evolution assuming the exponential potential. We then discuss the evolution of the Hubble expansion  parameter and the effective gravitational constant in the Friedmann equation.  The evolution of the Hubble parameter shows the tendency to fill the gap appearing in the Hubble tension problem.
In \S. V, we also present how to construct the potential when we know the evolution of the Hubble parameter and apply it to obtain the $\Lambda$CDM model.

In \S. VI, we analyze the density perturbations. We find the gravitational constant in the evolution equation of the density contrast is modified and becomes time-dependent.
We then give  the constraints on the parameters in the theories from observation.
The discussion and remarks follow in \S. VII.

We also present the rescaling property in this model in Appendix A, the overview of the original Cuscuton gravity theory with the construction of a potential when we know the evolution of the Hubble parameter in Appendix B, the detailed analysis of the cosmological  dynamics 
for the exponential potential in Appendix C, the analysis for the case with a vacuum energy in Appendix D, and some peculiarity in the vacuum case in Appendix E.

\allowdisplaybreaks

\begin{widetext}
\section{Cuscuta-Galileon Theory}
\label{Cuscuta-Galileon Theory}
We discuss  the Cuscuta-Galileon gravity, in which 
the minimum contribution of a Galileon-type scalar field  is included in the Cuscuton gravity theory.
The action is given by
\bea
S &=& \int d^4 x \sqrt{-g} \Big[ \frac{1}{2} \mpl^2 R + \alpha_2 \mpl^2 \sqrt{-X} + \alpha_3\mpl \ln \Big(-\frac{X}{\Lambda^4}\Big) \square \phi  - V(\phi)   +3 \alpha_3^2 X \Big] + S_M (g_{\mu\nu}, \psi_M) \,, \nn
\label{action0}
\ena
where $X$ is defined as 
\beann
X \equiv g^{\mu\nu}\partial_{\mu} \phi \partial_{\nu} \phi\,,
\enann
and $\alpha_2$ and $\alpha_3$ are dimensionless coupling constants, respectively, while 
$\Lambda$ is a cutoff-scale constant with mass dimension. 
This model is one of a special case of the extended Cuscuton gravity theory \cite{Iyonaga:2018vnu}; however, it is not the same as their application to dark energy \cite{Iyonaga:2020bmm}.
The original Cuscuton model is obtained by setting $\alpha_2=\mu^2/\mpl^2$ and $\alpha_3=0$.
This action has also found by covariantization of the minimally modified gravity \cite{Mukohyama:2019unx}.

Taking the variation of the above action with respect to the scalar field $\phi$ and the metric $g_{\mu\nu}$,
we find the following basic equations:
\beann
&&
\alpha_2\mpl^2 {1\over \sqrt{-X}}\left[\square \phi-{1\over 2X}\nabla X\cdot \nabla \phi\right]
+\alpha_3\mpl \left[-2\nabla  \left({\square \phi\over X}\right)\cdot \nabla  \phi
-2{\left(\square\phi\right)^2\over X}+\square\left(\ln (-X)\right)\right]-6\alpha_3^2 \square  \phi-V_{,\phi}=0\,,
\\
&&
\mpl^2 G_{\mu\nu}=T_{\mu\nu}-g_{\mu\nu}V+\alpha_2
\mpl^2 {1\over \sqrt{-X}}\left[-g_{\mu\nu} X +\partial_\mu\phi \partial_\nu\phi\right]
\\
&&~~~~~~~~
+\alpha_3\mpl \left[-2\partial_\mu\phi\partial_\nu\phi{\square \phi\over X}+{2\over X}\partial_{(\mu}X\partial_{\nu)}\phi-g_{\mu\nu} {1\over X}\left(
\nabla  X\cdot \nabla \phi \right)\right]+3\alpha_3^2\left(g_{\mu\nu} X-2\partial_\mu\partial_\nu \phi\right)\,.
\enann
Assuming the conservation of energy-momentum of matter field, i.e., 
$
\nabla^\nu T_{\mu\nu}=0\,,
$
and using the Bianchi identity $\nabla^\nu G_{\mu\nu} \equiv 0\,,$
we recover  the first equation for $\phi$ from the second Einstein equations.
Hence only the Einstein equations are independent  in the present model.
We do not have additional degrees of freedom in addition to the Einstein equations, 
The scalar field $\phi$ does not carry new degree of freedom just as the original Cuscuton.
We will prove it below.

Note that this model is completely different from the original one 
($\alpha_3=0$).
Because as we show in Appendix \ref{rescaling}, 
we can always set $\alpha_3=1$ without loss of generality, which means 
that the perturbation approach for the original theory  does not provide 
 an appropriate approximation  even for the case of $|\alpha_3|\ll 1$.
 However we shall keep $\alpha_3$ in the text in order to see the coupling dependence. 
Since the results for $\alpha_3<0$ can be obtained by the change of the sign of $\phi$,
we assume $\alpha_3\geq 0$ in this paper.

\subsection{Degrees of freedom}

According to the method in Refs. \cite{Tsujikawa:2014mba,Kase:2014yya,Kase:2014cwa} we use the $3+1$ decomposition metric and choose the unitary gauge:
\beann
ds^2 = - N^2 dt^2 + h_{ij} (dx^i + N^i dt)(dx^j + N^j dt) \,, \quad \phi = \phi(t) \,,
\enann
the action (\ref{action0}) can be written in the Arnowitt-Deser-Misner (ADM) form as 
\beann
S = \int dt d^3 x N \sqrt{h} \left[ \frac{1}{2}\mpl^2 \Big( {}^3 R + K^{ij} K_{ij} - K^2 \Big) + \left( \frac{\alpha_2 \mpl^2 |\dot\phi|}{N} - V(\phi) - \frac{3 \alpha_3^2 \dot\phi^2}{N^2} \right) + \left( -\frac{2 \alpha_3 \mpl |\dot\phi|}{N} + C_1 \right) K \right] \,. 
\enann
Note that in this section we will not consider contribution from the matter Lagrangian. ${}^3 R$ is the three-dimensional Ricci scalar, $K_{ij}$ is the extrinsic curvature, $K$ is the trace of $K_{ij}$, and $C_1$ is an integration constant. 

Following calculations in Ref. \cite{Lin:2014jga} since the scalar field is a function of time, the fundamental variables are only $N$, $N^i$, and $h^{ij}$ which are the lapse function, the shift vector, and the three-dimensional metric, respectively. Their conjugate momenta are
\beann
&&
\pi_N = \frac{\partial {\cal L}}{\partial \dot N} = 0 \,, \quad \pi_i = \frac{\partial {\cal L}}{\partial \dot N^i} = 0 \,, 
\\
&&
\pi^{ij} = \frac{\partial {\cal L}}{\partial \dot h^{ij}} = \frac{1}{2} \sqrt{h} \left( -\frac{2 \alpha_3 \mpl |\dot\phi|}{N} + C_1 \right) h^{ij} - \frac{1}{2} \mpl^2 \sqrt{h} \left(K h^{ij} - K^{ij}\right) \,. \nn
\enann
Thus the primary constraints are $\pi_N$ and $\pi_i$. Using the Legendre transformation, the Hamiltonian is given by 
\bea
H = \int d^3 x \left({\cal H} + N^i {\cal H}_i + \lambda_N \pi_N + \lambda^i \pi_i \right) \,, \label{Hamiltonian}
\ena
where 
\beann
{\cal H} &=& N \sqrt{h} \left[\frac{2}{\mpl^2} \left(\frac{\pi^{ij} \pi_{ij} }{h} - \frac{\pi^2}{2h}\right) - \frac{1}{2}\mpl^2 {}^3 R - \left(\frac{\alpha_2 \mpl^2 |\dot\phi|}{N} - V(\phi)\right) \right. \nn
&& \left. + \frac{\pi}{\mpl^2 \sqrt{h} }\left(-\frac{2\alpha_3 \mpl |\dot\phi|}{N} + C_1\right) - \frac{3}{4\mpl^2} \left(C^2_1 - \frac{4\alpha_3 \mpl C_1 |\dot\phi|}{N}\right)\right]  \nn
{\cal H}_i &=& - 2 h_{ik} D_j \pi^{kj} \,,
\enann
and $\lambda_N$ and $\lambda^i$ are Lagrange multipliers. 
\end{widetext}
The secondary constraints are given by
\beann
0 &=& \dot \pi_N = - \frac{\partial H}{\partial N} \approx - \frac{\partial {\cal H}}{\partial N} \equiv {\cal C} \,, \\
0 &=& \dot \pi_i = - \frac{\partial H}{\partial N^i} \approx {\cal H}_i \,.
\enann
The $\approx$ means equality when the constraints are imposed. However the momentum constraint is not a first-class constraint because one of the Poisson brackets with other constraints does not vanish. Therefore we introduce
\beann
\bar{\cal H}_i = {\cal H}_i + \pi_N \partial_i N \,.
\enann

On the constraint surface we find $\bar{\cal H}_i = {\cal H}_i ~(\approx 0)$ because of $\pi_N = 0$. Then we can consider $\bar{\cal H}_i$ as the momentum constraint. The Poisson brackets of constraints are (see definition of the Poisson bracket in Ref. \cite{Lin:2014jga,Panpanich:2021lsd})
\beann
&&
\{\pi_i (x), \pi_N (x^{\prime})\} = 0 \,, 
\\
&&
\{\pi_i (x), {\cal \bar H}_j (x^{\prime})\} =0 \,, 
\\
&&\{\pi_i (x), {\cal C} (x^{\prime})\} = 0 \,, 
\enann
\beann
&&
\{ {\cal \bar H}_i [f^i], \bar \pi_N [\varphi] \} = \int d^3 y \pi_N f^i \partial_i \varphi \approx 0 \,, \\
&&
\{ {\cal \bar H}_i [f^i], {\cal C} [\varphi] \}= \int d^3 y {\cal C} f^i \partial_i \varphi \approx 0 \,, 
\\
&&
\{\pi_N (x), {\cal C} (x^{\prime})\} = \frac{\partial^2 {\cal H}}{\partial N^2} \delta(x - x^{\prime}) \,, 
\enann
where we have used the smeared constraint forms which are defined as
\beann
&&
{\cal \bar H}_i [f^i] \equiv \int d^3 x f^i (x) {\cal \bar H}_i (x)  \\
&&
\bar \pi_N [\varphi] \equiv \int d^3 x \varphi (x) \pi_N (x)  \\
&&
{\cal C} [\varphi] \equiv \int d^3 x \varphi (x) {\cal C} (x) \,.
\enann
Since ${\cal H}$ in the Hamiltonian (\ref{Hamiltonian}) is a linear function of the lapse function, i.e. $\partial^2 {\cal H}/\partial N^2  = 0$, the last Poisson bracket is equal to zero. Consequently, all of constraints are the first-class constraints. 

We have $10$ variables which is equal to $20$ dimensions in phase space with $8$ first-class of constraints. Thus degrees of freedom of the theory can be calculated by
\bea
{\rm d.o.f.} &=& \frac{1}{2} \Big({\rm variables} \times 2 - {\rm 1st ~ class} \times 2 - {\rm 2nd ~ class}\Big)  \nn
&=& \frac{1}{2} \Big(10 \times 2 - 8 \times 2 -0 \Big)  \nn
&=& 2 \,.
\ena
As a result, the theory has $2$ degrees of freedom.

~\\

\begin{widetext}

\section{Dynamics of FLRW spacetime}
\label{FLRW spacetime}

We consider the flat Friedmann-Lema\^{\i}tre-Robertson-Walker (FLRW) metric and choose the unitary gauge as
\beann
ds^2 = - N(t)^2 dt^2 + a(t)^2 d\vect{x}^2 \,, \quad  \phi = \phi(t) \,.
\enann
Substituting the metric into the above action, and then varying with respect to $\phi$, $N$, and $a$, after setting $N = 1$ we find
\bea
&&
6\alpha_3 \mpl\left(3  H^2 +  \dot H\right) - V_{,\phi} - 3\alpha_2\mpl^2  H \epsilon_{\dot\phi} + 6 \alpha_3^2
 \left(3 H \dot\phi    + \ddot\phi\right)
= 0 \,, 
\label{EOM}
\\
&&
3 \mpl^2 H^2 - \rho- V(\phi) + 6 \alpha_3\mpl H \dot \phi   +3\alpha_3^2  \dot\phi^2= 0 \,,
\label{1stFriedmann} 
\\
&&
3\mpl^2  H^2 + 2 \mpl^2 \dot H + P- V(\phi) + \alpha_2 \mpl^2 \lvert \dot \phi \rvert    + \alpha_3
\left(2\mpl \ddot \phi - 3 \alpha_3 \dot\phi^2\right)
 = 0 \,, \label{2ndFriedmann} 
\ena
where $\epsilon_{\dot\phi}\equiv {\rm sgn} ({\dot\phi})$, and 
$\rho$ and $P$ are total matter density and pressure, respectively. 
Since  $\epsilon_{\dot\phi}$ changes the value discretely, 
we should assume that $\dot \phi\neq 0$ and 
it does not change the sign during the evolution of the universe.
We may have two branch solutions.
The solution with $\phi$=constant is incompatible with the timelike ansatz.
However we can discuss the limiting case as $\dot \phi\rightarrow 0$,
which will give two different solutions 
unless $H$ vanishes.

We may assume that matter components consist of perfect fluids such as matter and radiation, that is, 
\bea
\rho=\sum_{i} \rho_i\,, ~P=\sum_{i} P_i\,,~~{\rm with } ~P_i=w_i\rho_i\,,
\label{matter_fluid}
\ena
where $w_i$ describes the equation of state of $i$ th matter component.
The matter ($\rho_m, P_m$) and radiation  ($\rho_r, P_r$)  are given by
$w_m=0$ and $w_r={1\over 3}$, respectively.

As we show in \S.\ref{Cuscuta-Galileon Theory},
the first equation for the scalar field $\phi$ is derived from the Einstein equations.
In what follows, we rewrite the above basic equations to solve them.

Introducing new Hubble parameter by
\beann
\bar{H}\equiv H+\alpha_3\mpl^{-1}\dot \phi
\,,
\enann
we rewrite the above three equations of motion as
\bea
&&
6 \bar{H}^2 + 2  \dot{\bar{H}} -6\alpha_3\mpl^{-1}\bar{H}\dot{\phi}
+ \alpha_2 |\dot{\phi}|
- {1\over 3 \alpha_3\mpl}V_{,\phi} - {\alpha_2  \over \alpha_3}\mpl \bar{H} \epsilon_{\dot\phi}
=0 \,, \label{new_EOM2}
\\
&&3 \mpl^2 \bar{H}^2 =\rho+V(\phi)  \,,\label{new_1stFriedmann} \\
&&
3 \bar{H}^2 +  2\dot{\bar{H}} 
-6\alpha_3\mpl^{-1}\bar{H}\dot{\phi}+ \alpha_2 |\dot{\phi}| +\mpl^{-2}\left(P - V(\phi) \right)
=0 \,.
 \label{new_2ndFriedmann2} 
\ena
From Eqs. (\ref{new_EOM2}) and (\ref{new_2ndFriedmann2}), we find 
\beann
\bar{H}^2&=& {1\over 3\mpl^2 }\left(P- V(\phi) \right)
+ {1\over 9\alpha_3\mpl}V_{,\phi}+  {\alpha_2  \over 3\alpha_3}\mpl   \bar{H} \epsilon_{\dot\phi} \,.
 \label{Hubble2}
\enann
With Eq. (\ref{new_1stFriedmann}), we obtain the following equation
\bea
\bar{H} \epsilon_{\dot\phi}= {\alpha_3\over \alpha_2\mpl^3}\left[\rho
 -P +2V(\phi)- {\mpl\over 3\alpha_3}V_{,\phi}\right] \,, \label{newHsgn}
\ena
or
\bea
\bar{H}^2
= {\alpha_3^2\over \alpha_2^2\mpl^6}\left[\rho-P +2V(\phi)- {\mpl\over 3\alpha_3}V_{,\phi}\right]^2 \,.
\label{newHsgn2}
 \ena
From Eq. (\ref{new_1stFriedmann}) and Eq. (\ref{newHsgn2}), we obtain one constraint equation 
 \bea
{1\over 3 }\left(\rho+V(\phi)\right)=  {\alpha_3^2\over \alpha_2^2\mpl^4}\left[\rho-P +2V(\phi)- {\mpl\over 3\alpha_3}V_{,\phi}\right]^2 \,.
\label{constraint}
\ena
\end{widetext}

This constraint equation gives the relation between  the scalar field $\phi$ and matter and radiation, once we assume the potential $V(\phi)$.
The scalar field $\phi$ is no longer dynamical,
but it is fixed by matter fluid ($\rho\,,P$).

For the perfect fluids, we find the time evolution of their densities as
\beann
&&
\rho_i \propto a^{-3(1+w_i)}\,,~P_i=w_i\rho_i \,,
\enann
from the energy conservation equation.
Hence $\rho$ and $P$ are given by some known function of the $e$-folding number  $N\equiv \ln (a/a_0)$ as
$\rho(N)$ and $P(N)$, where $a_0$ is the present value of the scale factor.

Solving the  constraint equation (\ref{constraint}) for the scalar field $\phi$  in terms of the $e$-folding number  $N$, we find
\beann
\phi=\phi(N)\,.
\enann
Since 
\beann
\bar H=H+{\alpha_3\over \mpl} \dot\phi=HZ(N)\,,
\enann
where
\beann
Z(N)&\equiv& 1+{\alpha_3\over \mpl} {d\phi\over dN} \,,
\enann
we obtain the effective Friedmann equation from Eq. (\ref{new_1stFriedmann}) as
\bea
H^2= {1\over 3\mpl^2 Z^2(N)}
\left[\rho(N)+V(\phi(N))\right]
\,.\label{effective_Friedmann_equation}
\ena
This equation gives the solution of the scale factor,  $a=a(t)$. The prefactor 
$Z^{-2}$ modifies  the Friedmann equation from the general relativistic one.
Note that there is no kinetic term of a scalar field.

\section{Exponential potential}
\label{expo_potential}
In order to analyze the cosmological evolution, 
we have to give a concrete form of the potential $V(\phi)$.
Here we shall assume an exponential potential, 
\bea
V=\epsilon_V\mpl^4 \exp\left(\lambda \alpha_3 \mpl^{-1}\phi\right)
\label{exp_pot}
\,,
\ena
where $\lambda$ is a coupling constant. Without loss of generality, we can normalize the coefficient of the potential as $\epsilon_V=\pm 1$ because of rescaling of a scalar field $\phi$.

Assuming there exist matter and radiation  as 
matter components,
the constraint equation (\ref{constraint}) is 
 \beann
{1\over 3}\left(\rho_m + \rho_r+V\right)
= {\alpha_3^2\over \alpha_2^2\mpl^4}\left[\rho_m + {2\over 3}\rho_r+\left(2- {\lambda\over 3}\right)V\right]^2
 \,,
\enann
which is rewritten as
 \bea
 &&
 \left(2- {\lambda\over 3}\right)^2V^2+\left[2 \left(2- {\lambda\over 3}\right)\left(\rho_m+{2\over 3}\rho_r\right)
 -{\alpha_2^2\mpl^4 \over 3\alpha_3^2}\right]V
\nonumber \\
 &&
 ~~~+\left(\rho_m+{2\over 3}\rho_r\right)^2-{\alpha_2^2\mpl^4 \over 3\alpha_3^2}\left(\rho_m+\rho_r\right)=0
 \,.
 \label{potential_rho}
\ena
This must have a real solution for $V$.
If $\lambda=6$, we always have a simple solution
\beann
V=
{3\alpha_3^2\over \alpha_2^2\mpl^4} \left(\rho_m+{2\over 3}\rho_r\right)^2-\left(\rho_m+\rho_r\right)
\,.
\enann

For the case of $\lambda \neq 6$, we have a quadratic equation.
Before solving it, we shall take the limit of $a\rightarrow \infty$  (or 
equivalently $\rho_m\,,\rho_r \rightarrow 0
 $).
Eq. (\ref{potential_rho}) gives 
\beann
V\left(V-V_\infty\right)=0\,,
\enann
where
\bea
V_\infty\equiv {3\alpha_2^2\over (\lambda-6)^2\alpha_3^2}\mpl^4\,.
\ena

We then normalize the variables and parameters by $V_\infty$, which are described by those with a tilde.
The quadratic equation for $V$ is now
 \bea
 &&
 \tilde V^2-\left[{2\over \lambda-6}\left(3\tilde \rho_m+2\tilde \rho_r\right)+1\right]\tilde V
 \nonumber \\
 &&~~
 +{1\over \left(\lambda-6\right)^2}
 \left(3\tilde \rho_m+2\tilde \rho_r\right)^2-\left(\tilde \rho_m+\tilde \rho_r\right)=0
\label{potential_rho2}
 \,,
 \ena
 where
\beann
\tilde V \equiv {V\over V_\infty}
\,,~
\tilde \rho_m \equiv {\rho_m\over V_\infty}
\,,~
{\rm and}~~
\tilde \rho_r \equiv {\rho_r\over V_\infty}
\,.
\enann

In order to have a real solution for $\tilde V$, 
 the following condition should be satisfied:
\beann
D
\equiv 
1+{4\over \lambda-6}
\left[(\lambda-3)\tilde \rho_m+(\lambda-4)\tilde \rho_r\right] 
\geq 0 \,.
\enann
This condition gives the constraint on $\rho_m$ and $\rho_r$.
We can classify the possible cases by the exponent $\lambda$ of the potential.
We summarize the classification in Table \ref{table1}, in which we show
the range of $\tilde \rho_m$ and $\tilde \rho_r$ for existence  of 
a real solution  $\tilde V$.

\begin{table}[h]
\begin{tabular}{ lc| c|}
&exponent& existence range
\\
  \hline  \hline 
(a) &$\lambda>6$&$0\leq \tilde\rho_m,\tilde\rho_r<\infty $\\
  \hline  
(b) &$\lambda= 6$&$0\leq \tilde\rho_m,\tilde\rho_r<\infty $\\
  \hline  
(c) &$4< \lambda<6$ &$ (\lambda-3)\tilde \rho_m+(\lambda-4)\tilde\rho_r\leq{1\over 4}(6-\lambda) $ \\
   \hline
(d) &$\lambda=4$ &$  \tilde \rho_m \leq{1\over 2}$   \\
  \hline
(e) & $3<\lambda<4$&$ (\lambda-3) \tilde\rho_m\leq (4-\lambda)\tilde \rho_r +{1\over 4}(6-\lambda)  $ \\
  \hline  
(f) &$\lambda=3$ &$0\leq \tilde\rho_m,\tilde\rho_r<\infty $ \\
  \hline 
(g) &$0<\lambda<3$ &$0\leq \tilde\rho_m,\tilde\rho_r<\infty $ \\
  \hline 
(h) &$\lambda<0$ &$0\leq \tilde\rho_m,\tilde\rho_r<\infty $ \\
  \hline 
\end{tabular}
\caption{The existence range of $\tilde\rho_m$ and $\tilde\rho_r$ for the solution of E. (\ref{potential_rho2}) for $\tilde V$. 
For $\lambda\geq 6$ or $\lambda\leq 3$ ($\lambda\neq 0$), we find the full range of the densities. 
In the case of $4\leq \lambda<6$, there exists some upper bound on  
densities (or lower bound for a scale factor).
 For the case of $3<\lambda<4$, depending on the parameters,
 there are two possibilities (see the detail in the text and Appendix). }
\label{table1}
\end{table}

For $\lambda\geq 6$ or $\lambda\leq 3$ ($\lambda\neq 0$), we find the full range of the densities. 
In the case of $4\leq \lambda<6$, there exists some upper bound on  
densities (or lower bound for a scale factor).
 For the case of $3<\lambda<4$, depending on the parameters,
 there are two possibilities: Either  the full range of the densities is possible 
or 
two separated finite ranges of the scale factor are possible, i.e., $a\leq a_1$ or $a_2
\leq a$\, $(a_2<a_1)$. 
The latter case happens either when $\lambda$ is close to 4 or matter density is large enough.

The solution $\phi_\pm$ is given by
\bea
&&
\phi_\pm={\mpl\over \lambda\alpha_3}
\ln \left[{3\alpha_2^2\over \epsilon_V(\lambda-6)^2\alpha_3^2}\tilde V_\pm \right] \,,
\label{sol_phi_exp}
\\
\nonumber
&&
~
\ena
where
\beann
\tilde V_{\pm}&\equiv&\tilde V(\phi_\pm)=
{1\over 2}\left[
1 +{2\over \lambda-6}\left(3\tilde \rho_m+2\tilde \rho_r\right)
\pm \sqrt{D}\right]
\,.
\enann
We call them $\pm$ branches, respectively.
In order to exist the real solution, we have the constraint such that
\beann
\epsilon_V\tilde V_\pm(\tilde \rho_m,\tilde \rho_r; \lambda) \geq 0
\,,
\enann
which means that the potential is positive definite ($\epsilon_V=1$) for the case of 
$\tilde V_\pm(\tilde \rho_m,\tilde \rho_r; \lambda) >0$,
otherwise it is negative definite ($\epsilon_V=-1$).
$\epsilon_{\dot\phi} $ is determined from Eq. (\ref{sol_phi_exp}).
For example, for $\lambda>6$, $\phi_+$ decreases as $a$ increases (or densities decrease), which gives $\epsilon_{\dot\phi}=-1$.

\begin{widetext}

In order to derive the effective Friedmann equation
(\ref{effective_Friedmann_equation}), 
we have to evaluate the prefactor 
$Z^{-2}$.
Using the relation 
\beann
{\alpha_3\over \mpl}{d\phi_\pm \over d\ln a}
&=&{1\over \lambda}{d \ln V_\pm \over d\ln a}
=
{1\over \lambda}\left[{\partial \ln \tilde  V_\pm\over \partial\tilde  \rho_m}
{d \tilde \rho_m\over d\ln a}
+{\partial \ln \tilde V_\pm \over \partial\tilde  \rho_r}
{d \tilde \rho_r\over d\ln a}
\right]
=
-{1\over \lambda}\left[3{\tilde \rho_m\over\tilde  V_\pm}  {\partial \tilde V_\pm\over \partial\tilde \rho_m}
+4{\tilde \rho_r\over\tilde  V_\pm}
 {\partial \tilde  V_\pm\over \partial\tilde \rho_r}
\right]\,,
\enann
we find
\beann
&&
Z_\pm=1+{\alpha_3\over \mpl}{d\phi_\pm\over d\ln a}
={F_\pm(\tilde \rho_m\,,\tilde \rho_r; \lambda)
\over
S_\pm (\tilde \rho_m, \tilde \rho_r; \lambda) \sqrt{D}} \,,
\enann
where
\beann
&&
S_\pm (\tilde \rho_m, \tilde \rho_r; \lambda)\equiv
2\tilde  V_\pm=
1 +{2\over \lambda-6}\left(3\tilde \rho_m+2\tilde \rho_r\right)
\pm \sqrt{D} \,,
\\
&&
F_\pm(\tilde \rho_m\,,\tilde \rho_r; \lambda)\equiv
\left[1+{2\over \lambda(\lambda-6)}\left(3(\lambda-3)\tilde \rho_m+2(\lambda-4)\tilde \rho_r\right)\right]\sqrt{D}
\pm\left[1
+{2\over \lambda(\lambda-6)}\left((\lambda-3)(2\lambda-3)\tilde \rho_m+2(\lambda-2)(\lambda-4)\tilde \rho_r\right)\right] \,.
\enann

Since 
\beann
\rho_m+\rho_r+V(\phi_\pm)=\rho_m+\rho_r+V_\pm=
{V_\infty\over 2}R_\pm(\tilde \rho_m, \tilde \rho_r; \lambda)
\,,
\enann
where
\beann
R_\pm(\tilde \rho_m, \tilde \rho_r; \lambda)
\equiv 1 +{2\over \lambda-6}\left((\lambda-3)\tilde \rho_m+(\lambda-4)\tilde \rho_r\right)
\pm \sqrt{D}
\,,
\enann
we obtain the effective  Friedmann equation as
\bea
 H^2={1\over 3\mpl^2}{V_\infty D(\tilde \rho_m, \tilde \rho_r; \lambda) S_\pm^2 (\tilde \rho_m, \tilde \rho_r; \lambda) 
R_\pm (\tilde \rho_m, \tilde \rho_r; \lambda) \over 
2 F_\pm^2(\tilde \rho_m\,,\tilde \rho_r; \lambda)}
\,.~~~~~~~~
\label{effective_Friedmann}
\ena

\end{widetext}

~~
\\[-1.5cm]

\subsection{Two limiting stages}
We first consider two limiting stages ($a\rightarrow \infty$ and 
$a\rightarrow 0$), assuming their existence.
Those correspond to $\rho_m, \rho_r\rightarrow 0$ and $\rho_m, \rho_r
\rightarrow \infty$, respectively.

\subsubsection{$a\rightarrow \infty \,(\rho_m, \rho_r\rightarrow 0)$}
In this limit, the potentials for 
two branch solutions  are approximated as
\beann
\tilde V(\phi_+)=\tilde V_+&\approx &1+{1\over \lambda-6}\left[\lambda\tilde \rho_m+(\lambda-2)\tilde \rho_r\right] 
\\
\tilde V(\phi_-)=\tilde V_-&\approx & -(\tilde \rho_m+\tilde \rho_r)+{\left[(\lambda-3)\tilde \rho_m+(\lambda-4)\tilde \rho_r\right]^2\over (\lambda-6)^2} \,.
\enann
We find $\epsilon_V=+1$ for $+$ branch
($\tilde V_+>0$), while $\epsilon_V=-1$  for $-$ branch ($\tilde V_-<0$).

For the $+$ branch solution $\tilde V_+$,
 we find
\beann
\mpl^2 H^2\approx {V_\infty\over 3} \,,
\enann
which gives the de-Sitter type accelerating universe as
\bea
a(t)\propto \exp\left[H_\infty t\right] \,,
\ena
where 
\beann
H_\infty\equiv \mpl^{-1} \sqrt{V_\infty\over 3}={|\alpha_2|\over (\lambda-6)\alpha_3}\mpl
\,.
\enann
In order to explain the present acceleration of the universe in this model, 
we have a strong constraint on the coupling constant as
\beann
{|\alpha_2|\over (\lambda-6)\alpha_3}\sim O\left(10^{-60}\right)\ll 1\,.
\enann

The scalar field approaches 
as
\beann
\phi\rightarrow \phi_\infty\equiv {\mpl\over \lambda \alpha_3}\ln \left[{3\alpha_2^2\over (\lambda-6)^2\alpha_3^2}\right]
\,.
\enann

For the other branch solution $\phi_-$,
we find
\beann
\mpl^2 H^2={1\over 3Z^2}\left(\rho_m+\rho_r+V_-\right)
\approx 
{\lambda^2 \over 3 (\lambda-6)^2}V_\infty
\rho_m^2\propto {1\over a^6} \,,
\enann
because  $\rho_m\gg \rho_r$ as $a\rightarrow \infty$.
It gives the asymptotic behaviour
as
\beann
a(t)\propto t^{1/3}
\,,
\enann
which is the expansion law for the stiff matter ($P=\rho$) 
in general relativity (GR), although 
matter density dominates the universe.
The scalar field approaches 
as
\beann
\phi\rightarrow -\infty\,.
\enann

\subsubsection{ $a\rightarrow 0 ~(\rho_m, \rho_r \rightarrow \infty)$}

The asymptotic behaviours of the two branch solutions ($V_\pm$) and  the Friedmann equation 
 become the same forms as
\beann
\tilde V_\pm\approx {1\over \lambda-6}\left(3\tilde \rho_m+2\tilde \rho_r\right) \,,
\enann
and 
\beann
\mpl^2 H^2 &\approx &
{\lambda^2\over 3(\lambda-6)}{\left(3\rho_m+2\rho_r\right)^2
\left[(\lambda-3)\rho_m+(\lambda-4)\rho_r\right]
\over \left[3(\lambda-3)\rho_m+2(\lambda-4)\rho_r\right]^2} \,.
\enann

If $\lambda\neq 3, 4 $, 
\beann
\mpl^2 H^2 &\approx&
\left\{
\begin{array}{cccl}
{\lambda^2\over 3(\lambda-3)(\lambda-6)}\rho_m
&{\rm for} &\rho_m\gg\rho_r&{\rm (MD)}
\\
{\lambda^2\over 3(\lambda-4)(\lambda-6)}\rho_r
&{\rm for} &\rho_m\ll\rho_r&{\rm (RD)}
\\
\end{array}
\right. \,,
\enann
where MD and RD denote matter dominant stage and radiation dominant stage, respectively.

This gives
\beann
a(t)\propto 
\left\{
\begin{array}{cccl}
t^{2\over 3}
&{\rm for} &\rho_m\gg\rho_r&{\rm (MD)}
\\
t^{1\over 2}
&{\rm for} &\rho_m\ll\rho_r&{\rm (RD)}
\\
\end{array}
\right.
\,,
\enann
which is the same as the evolution history in the standard big-bang model.
However the effective gravitational constant in the Friedmann equation
$G_{\rm F}$  is 
different from the Newtonian gravitational constant $G_{\rm N}\equiv
(8\pi\mpl^2)^{-1}$.
Note that the scalar field approaches in this limit as
\beann
\phi_\pm \rightarrow \infty\,,
\enann
for both branches.

$G_{\rm F}$   shows a gap between the values at 
radiation dominant stage 
and at  the matter dominant stage.
In fact, we find 
\beann
G_{\rm F}=
\left\{
\begin{array}{cccl}
{\lambda^2\over (\lambda-3)(\lambda-6)}G_{\rm N}
&{\rm for} &\rho_m\gg\rho_r&{\rm (MD)}
\\
{\lambda^2\over (\lambda-4)(\lambda-6)}G_{\rm N}
&{\rm for} &\rho_m\ll\rho_r&{\rm (RD)}
\\
\end{array}
\right. \,.
\enann

One may wonder what happens if  $3\leq \lambda\leq 6$, when $G_{\rm F}<0$.
As we show in Appendix \ref{exponential_potential}, in such a case,  there is no limit of $a\rightarrow 0$.
The scale factor $a$ is bounded from below, that is $ a\geq a_{\rm min} (>0)$.

In the cases of $\lambda=3$ and $\lambda=4$, 
we find  strange behaviours in the Friedmann equation as follows:
For $\lambda=3$, 
\beann
\mpl^2 H^2 &\approx&
\left\{
\begin{array}{cccl}
{9\over 4}{\rho_m^2\over \rho_r}
&{\rm for} &\rho_m\gg\rho_r&{\rm (MD)}
\\
\rho_r
&{\rm for} &\rho_m\ll\rho_r&{\rm (RD)}
\\
\end{array}
\right.
\,,
\enann
which expansion law becomes 
\beann
a(t)  &\propto&
\left\{
\begin{array}{cccl}
t
&{\rm for} &\rho_m\gg\rho_r&{\rm (MD)}
\\
t^{1\over 2}&{\rm for}&\rho_m\ll\rho_r&{\rm (RD)}
\\
\end{array}
\right.
\,.
\enann

On the other hand, for  $\lambda=4$, 
there exists no solution in this limit.

\subsection{ Whole history}

In the two liming stages, we may find 
an appropriate evolution of the universe, i.e., 
radiation/matter dominance in the early stage  
($a\rightarrow 0$), and de Sitter expansion for $+$ branch 
in the early stage 
($a\rightarrow \infty$).
However the above two limiting stages can be disconnected if 
there exists some finite scale factor at which
 the Hubble parameter $H$ vanishes or diverges, or $H^2$ becomes negative.
 It may happen
when one of the following conditions is satisfied
\beann
&{\rm (i)}~ &
D(\tilde \rho_m\,,\tilde \rho_r; \lambda)\leq 0 \,,
\\
&{\rm(ii)~ } &
S_\pm(\tilde \rho_m\,,\tilde \rho_r; \lambda)=0\,,
\\&{\rm(iii)~ } &
R_\pm(\tilde \rho_m\,,\tilde \rho_r; \lambda)
\leq 0\,,
\\
&{\rm (iv)~ } &
F_\pm(\tilde \rho_m\,,\tilde \rho_r; \lambda)=0\,.
\enann

 In fact $H$ vanishes when $S_\pm(\tilde \rho_m\,,\tilde \rho_r; \lambda)=0$, while 
 it diverges when $F_\pm(\tilde \rho_m\,,\tilde \rho_r; \lambda)=0$.
 In those cases, the above two limits are disconnected at that point.
 On the other hand,  when $D(\tilde \rho_m\,,\tilde \rho_r; \lambda)<0$ or 
$R_\pm(\tilde \rho_m\,,\tilde \rho_r; \lambda)<0$, 
no solution exists in such a range of densities $\tilde \rho_m\,, \tilde \rho_r$ 
(or a scale factor $a$).

In what follows, we just discuss one simple case ($\lambda>6$).
For the other cases, we show them in Appendix \ref{exponential_potential}.

\subsubsection{Exponential potential with $ \lambda> 6$}
\label{exp_pot_lam>6}
In this case, we find $D>0$, which guarantees the solution exists for full range of densities, i.e. 
$0\leq \rho_m,\rho_r<\infty $.
For the $\phi_-$ branch, 
there exists one point where $H$ vanishes, that is,
it happens when 
\beann
(3\tilde \rho_m+2\tilde \rho_r)^2=(\lambda-6)^2(\tilde \rho_m+\tilde \rho_r)\,.
\enann
which is obtained from the condition (ii).
We find the corresponding scale factor $a_{\rm cr}$  as
\beann
\rho_r(a_{\rm cr})={(\lambda-6)^2\over 4}V_\infty&~~{\rm if~it~happens~in}&~~{\rm RD} 
\\
\rho_m(a_{\rm cr})={(\lambda-6)^2\over 9}V_\infty&~~{\rm if~it~happens~in }&~~{\rm MD}
\,.
\enann
Since we find 
\beann
H^2\propto \left(a-a_{\rm cr}\right)^2
\,,
\enann
near $a=a_{\rm cr}$, the universe approaches $a_{\rm cr}$ exponentially with respect time
as
\beann
a(t)\approx a_{\rm cr}\mp a_* \exp (\mp K_* t)&~~~{\rm as}&~~~t\rightarrow \pm \infty
\,,
\enann
where $a_*$ and $K_*$ are positive constants.
As a result, we have two histories of the universe ($a_1(t)$ and $a_2(t)$) as
\beann
a_1(t)  &\propto&
\left\{
\begin{array}{cccccl}
&t^{1\over 2} &
&~~~{\rm in} &t\rightarrow 0
\\
 &({\rm RD})&
&&
\\
&a_{\rm cr}&&~~~{\rm as}&t\rightarrow \infty
\\
\end{array}
\right. \,,
\\[1em]
&&{\rm or}
\\[1em]
a_1(t)  &\propto&
\left\{
\begin{array}{cccccl}
t^{1\over 2} &\rightarrow& t^{2\over 3} 
&~~~{\rm in} &{\rm the~early ~stage}
\\
 ({\rm RD})&& ({\rm 
MD})
&&
\\
&a_{\rm cr}&&~~~{\rm as}&t\rightarrow \infty
\\
\end{array}
\right. \,,
\enann
and
\beann
a_2(t)  &\propto&
\left\{
\begin{array}{ccl}
a_{\rm cr}
&{\rm as} &t\rightarrow -\infty
\\
t^{1\over 3}&{\rm as}&t\rightarrow \infty
\\
\end{array}
\right.
\,.
\enann

For the $+$ branch, 
both denominator and numerator in the right hand side of the Friedmann equation
(\ref{effective_Friedmann}) do not vanish for any values of $\rho_m, \rho_r$.
Hence the above two limits are connected.
We find radiation dominant era and matter dominant era 
 in the early stage of the universe, which is followed by de Sitter 
 accelerating expansion.
 \beann
 a(t)  &\propto&
\left\{
\begin{array}{cccccl}
t^{1\over 2} &\rightarrow& t^{2\over 3} 
&~~~{\rm in} &{\rm the~early ~stage}
\\
 ({\rm RD})&& ({\rm 
MD})
&&
\\
&\exp(H_\infty t)
&&~~~{\rm as}&t\rightarrow \infty
\\
\end{array}
\right. \,.
\enann

\subsubsection{Summary of exponential potential}
\label{summary_exppot}

\begin{figure}[h]
	\includegraphics[width=6cm]{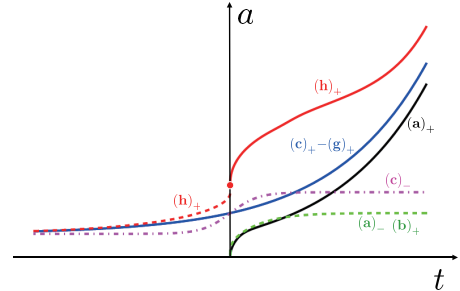} 
	\caption{The schematic evolution curves of the universe with positive exponential potential.  (a), (b), $\cdots,$ (h) correspond to the classification in Table 
	\ref{table2} and the suffixes $\pm$ denote the branches. }
	\label{schematic evolution1}
\end{figure}

Here we summarize the results on the cosmic evolution in Tables \ref{table2}, 
\ref{table3} and Figs. \ref{schematic evolution1}, \ref{schematic evolution2}. 
The details for the case of  $\lambda\leq 6$ are given  in Appendix 
\ref{exponential_potential}.

\begin{table}[h]
\begin{tabular}{| lc| c|c|}
  \hline
  &exponent~~~~& $+$ branch & $-$ branch
\\
  \hline  \hline 
  (a) &$\lambda>6$ 
  &RD/MD$\rightarrow$ dS&
  RD/MD $\rightarrow$M[$a_{\rm cr}$]
  \\
  \hline  
(b) &$\lambda=6$ 
 & \multicolumn{2}{c|}{P[1/4]$\rightarrow$ M$[ a_{\rm cr}]$  when $ \alpha_2/\alpha_3\ll O(1)$}
 \\
  \hline  
(c) &$4<\lambda<6$ 
& M[$a_{\rm min}$] $\rightarrow$ dS  &M[$a_{\rm min}$] $\rightarrow$M[$a_{\rm cr}$] \\
   \hline
(d) &$\lambda=4$
 & M[$a_{\rm cr}$] $\rightarrow$ dS &NA\\
  \hline
(e) &$3<\lambda<4$ 
& M[$a_{\rm cr}$] $\rightarrow$ dS& NA\\
  \hline  
(f) &$\lambda=3$ 
&M[$a_{\rm cr}$] $\rightarrow$ dS& NA\\
  \hline 
  (g) &$0<\lambda<3$ 
  &M[$a_{\rm cr}$] $\rightarrow$ dS&  NA \\
  \hline 
 (h) &$\lambda<0$ 
 &M $[a_{\rm cr}^{(S)}]$ $\rightarrow $ S $[a_{\rm cr}^{(F)}]$&
NA  \\
& &
S $[a_{\rm cr}^{(F)}]$ $\rightarrow $ dS&
 \\
\hline
\end{tabular}
\caption{
The classification of 
cosmic evolution of the universe with the positive exponential potential ($\epsilon_V=1$).
RD/MD  denotes the Friedmann universe of radiation dominant stage, possibly  followed by matter dominant stage.
dS means de Sitter accelerating universe, while P[$p$] gives the  power-law expanding universe
 with the power-exponent $p$ 
 ($a\propto t^p$).  M[$a$] shows  
Minkowski spacetime with the scale factor $a$, while S[a] means a singularity at finite scale factor $a$.
}
\label{table2}
\end{table}

\begin{table}[h]
\begin{tabular}{| lc| c|c|}
  \hline
  &exponent~~~~& $+$ branch & $-$ branch
\\
  \hline  \hline 
  (a) &$\lambda>6$ &NA&M[$a_{\rm cr}$]$\rightarrow $  P[1/3] ]\\
  \hline  
(b) &$\lambda=6$ &\multicolumn{2}{c|}{P[1/4]$\rightarrow$ P[1/3] ~~~
when $ \alpha_2/\alpha_3\gsim O(1)$}\\ 
 & &\multicolumn{2}{c|}{M$[ a_{\rm cr}]$
 $\rightarrow $ P[1/3] ~~~~ when $ \alpha_2/\alpha_3\ll O(1)$}\\
  \hline  
(c) &$4<\lambda<6$ & NA  &M[$a_{\rm cr}$]$\rightarrow $  P[1/3] \\
   \hline
(d) &$\lambda=4$ & M[$a_{\rm min}$] $\rightarrow$M[$a_{\rm cr}$]  &M[$a_{\rm min}$] $\rightarrow $  P[1/3] \\
  \hline
(e) &$3<\lambda<4$ & M[$a_{\rm min}$] $\rightarrow$M[$a_{\rm cr}$] & M[$a_{\rm min}$]$\rightarrow $  P[1/3] \\
  \hline  
(f) &$\lambda=3$ &RD/MD $\rightarrow$M[$a_{\rm cr}$] & RD/MD $\rightarrow $  P[1/3] \\
  \hline 
  (g) &$0<\lambda<3$ &RD/MD  $\rightarrow$M[$a_{\rm cr}$] &  RD/MD $\rightarrow $  P[1/3]  \\
  \hline 
 (h) &$\lambda<0$ &RD/MD $\rightarrow $M$[a_{\rm cr}^{(S)}]$
 &
RD/MD $\rightarrow $  P[1/3]  \\ 
   \hline 
\end{tabular}
\caption{
The classification of 
cosmic evolution of the universe with the negative exponential potential ($\epsilon_V=-1$).
The notations are the same as those in Table \ref{table2}.
}
\label{table3}
\end{table}

\begin{figure}[h]
	\includegraphics[width=6cm]{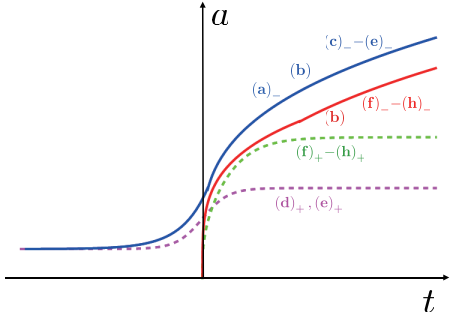} 
	\caption{The schematic evolution curves of the universe with negative exponential potential
	classified in Table \ref{table3}.
	The notations are the same as those in Fig. \ref{schematic evolution1}.}
	\label{schematic evolution2}
\end{figure}

As we show in the tables and schematic figures, the acceleration of the universe is obtained only for the $+$ branch solutions with a positive definite potential ($\epsilon_V=1$).
For the case with $\lambda<6$, 
we may not have radiation/matter dominant era in the early stage,
which is inconsistent with the big-bang nucleosynthesis.

\subsection{Gravitational ``constant'' in effective Friedmann equation and Hubble constant}

Since we are interested in the accelerating universe, 
we discuss the detail of the cosmological evolution 
for $+$ branch.

Using the redshift $z$, which is defined by $1+z=a_0/a$, 
the densities of matter and radiation are given by
\beann
\rho_m=3\Omega_{m,0}\mpl^2
H_0^2\left(1+z\right)^3\,,~~\rho_r=3\Omega_{r,0}\mpl^2
H_0^2\left(1+z\right)^4
\label{densities}
\,.
\enann

We then have 
\bea
\tilde \rho_m={\Omega_{m,0}\over \Omega_{\Lambda,0}}\left(1+z\right)^3\,,~~\tilde \rho_r={\Omega_{r,0}\over \Omega_{\Lambda,0}}\left(1+z\right)^4
\label{scaled_densities}
\,,
\ena
where 
\beann
 \Omega_{\Lambda,0}\equiv {V_\infty\over 3\mpl^2 H_0^2}
 \,.
\enann
Note that the $H_0$ here is based on the $\Lambda$CDM model. It is not the present value of the Hubble parameter in our model.

Inserting Eq. (\ref{scaled_densities}) into the Friedmann equation (\ref{effective_Friedmann}),
we find the Hubble parameter $H$ in terms of the redshift $z$.

\begin{figure}[h]
	\includegraphics[width=7cm]{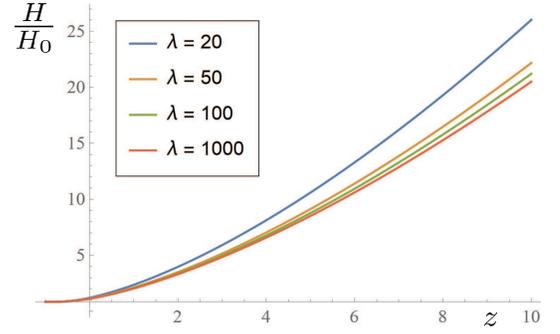} 
	\caption{Evolutions of $H$  in terms of the redshift $z$ (left figure).
	We set $\lambda=20, 50,  100\,, \Omega_{\Lambda,0}=0.7\,, \Omega_{m,0}=0.3\,,$ and $ \Omega_{r,0}=0.0001
	$.}
	\label{H}
\end{figure}

We show the result in Fig. \ref{H}

We rewrite the Friedmann equation (\ref{effective_Friedmann}) as follows:
\beann
H^2={8\pi G_{\rm F}(z)\over 3}
\left(\rho_m+\rho_r+V_\infty \right)
\,,
\enann
where 
$G_{\rm F}(z)$ is defined by
\beann
G_{\rm F}(z)= {1\over 16\pi \mpl^2}
{D(\tilde \rho_m,\tilde \rho_r\,;\lambda) S_+^2(\tilde \rho_m,\tilde \rho_r\,;\lambda) R_+(\tilde \rho_m,\tilde \rho_r\,;\lambda) 
\over 
(1+\tilde \rho_m+\tilde \rho_r) F_+^2(\tilde \rho_m,\tilde \rho_r\,;\lambda) }
\,.
\enann
If $G_{\rm F}=G_{\rm N}$, it gives the Friedmann equation in general relativity. Hence we can interpret  the effect on the
Friedmann equation by the Cuscuton $\phi$ 
as modification of the gravitational ``constant'' $G_{\rm F}$, which depends on $z$.
The asymptotic behaviour of $G_{\rm F}$ is given as
\bea
G_{\rm F}(z)\approx \left\{
\begin{array}{clll}
{\lambda^2\over (\lambda-4)(\lambda-6)}G_{\rm N}
&{\rm in} &{\rm RD}&(\rho_r\gg \rho_m\,, V_\infty )
\\
{\lambda^2\over (\lambda-3)(\lambda-6)}G_{\rm N}
&{\rm in} &{\rm MD}&(\rho_m\gg\rho_r\,,V_\infty )
\\
G_{\rm N}
&{\rm in} &{\rm DED}&(V_\infty\gg \rho_m\,,\rho_r)
\\
\end{array}
\right. \,,
~~~~~
\label{GF}
\ena
where
DED denotes dark energy dominant stage.

\begin{figure}[h]
	\includegraphics[width=7cm]{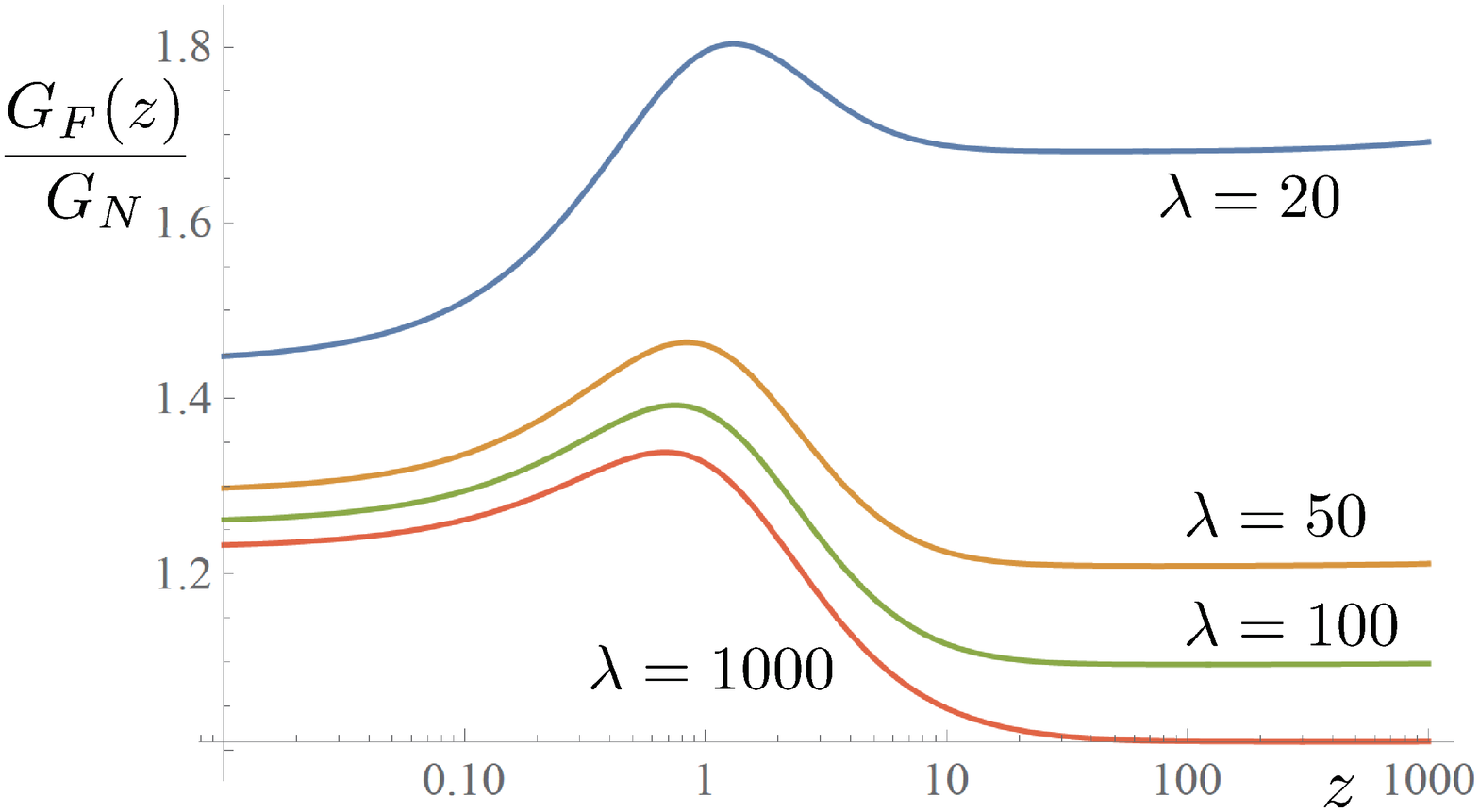}
	\caption{Evolutions of $G_{\rm F}$ in terms of the redshift $z$.
	We choose the same parameter values as those in Fig. \ref{H}. }
	\label{GFvsGN}
\end{figure}

We 
show some example of time evolution of $G_{\rm F}$ in Fig. \ref{GFvsGN}.

Since this gravitational ``constant'' $G_{\rm F}$ depends on time and it deviates 
from $G_{\rm N}$, we have the observational constraints by the big-bang nucleosynthesis
\cite{Alvey:2019ctk} such that
\bea
{G_{\rm BBN}\over G_{\rm N}}=0.99\begin{array}{c}
+0.06
\\
-0.05
\\
\end{array}
\,.
\ena
In the present model, $G_{\rm F}$ in the radiation dominant era is given by
Eq. (\ref{GF}), which gives the constraint on $\lambda$ as
\beann
\lambda\gsim 208\,.
\enann

\begin{figure}[h]
		\includegraphics[width=7cm]{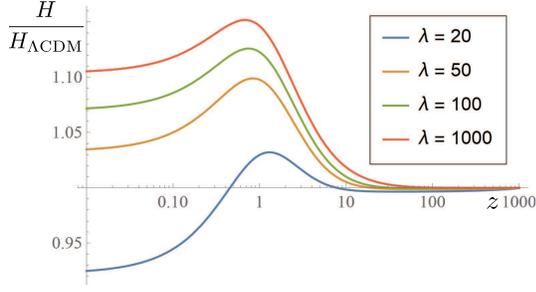} 
	\caption{Comparison with  $\Lambda$CDM model.
	The parameters are chosen as the same as Fig. \ref{H} }
	\label{HvsLCDM}
\end{figure}

We now present the comparison with  the $\Lambda$CDM model.
In Fig. \ref{HvsLCDM}, 
we present the evolution of the ratio of our Hubble expansion parameter to that in  the $\Lambda$CDM model,
$H/H_{\Lambda{\rm CDM}}(z)$, which is 
  normalized at $z=1100$ , i.e., 
$H/H_{\Lambda{\rm CDM}}(z=1100)=1$.

This figure shows 
that for $\lambda>200$, the Hubble expansion rate at  $z\leq 1$ is about 10\% larger 
than the value of  the $\Lambda$CDM model, which
 tendency might explain the Hubble tension
 \cite{Planck:2013pxb, Planck:2018vyg, Riess:2019cxk, Riess:2020fzl, DiValentino:2021izs}.
We shall discuss about it in \S. \ref{conclusion}.

\section{Construction of Appropriate Potential}
\label{construction}

Although the exponential potential may provide the interesting feature in the 
Cuscuta-Galileon gravity theory, it may not explain the observational data precisely.
Hence we shall discuss how to construct an appropriate potential $V(\phi)$ in our present 
model when the better evolution of the Hubble parameter is known.
 Once we can phenomenologically construct an appropriate potential from observational data, 
we might be able to find a fundamental theory behind it.

The basic equations are
\bea
&& HZ
\epsilon_{\dot\phi}=
{\alpha_3\over \alpha_2 \mpl^3}\left(\rho-P+2V-{\mpl\over 3\alpha_3}V_{,\phi} 
\right) \,,
\label{basic_eq1}
\\
&&
H^2Z^2={1\over 3\mpl^2}(\rho+V)
\,.
\label{basic_eq2}
\ena

The $e$-folding number $N\equiv \ln(a/a_0)$  measured from the present time
 is related to the redshift $z$ as
\beann
N=-\ln (1+z)\,.
\enann

Since 
\beann
{dV\over dN}=V_{,\phi} {d\phi\over dN}\,,
\enann
using the above basic equations, we find
\beann
&&
{d\over dN}\left[3\mpl^2 H^2\left(1+{\alpha_3\over \mpl}{d\phi\over dN}\right)^2-\rho\right]
\\
&&
~~
= {d\phi\over dN}\left[{\alpha_2\mpl^2 \over \alpha_3}
H\left(1+{\alpha_3\over \mpl}{d\phi\over dN}\right)
\epsilon_{\dot\phi}-(\rho-P)-2V\right] \,.
\enann

Eliminating $V$ and using the energy conservation 
\beann
{d\rho\over dN}+3(P+\rho)=0\,,
\enann
we obtain the second-order differential equation for $\phi$, 
which can be rewritten as
\bea
{dZ\over dN}+Q_1(N) Z-3Z^2=Q_2(N) \,,
\label{eq_Z}
\ena
where
\bea
Z&\equiv& 1+{\alpha_3\over \mpl}{d\phi\over dN} 
\label{def_Z}
\\
Q_1&\equiv&\left(3+{1\over H}{dH\over dN}+{\alpha_2\mpl\epsilon_{\dot\phi}\over 2\alpha_3 H}\right) 
\\
Q_2&\equiv&-{1\over 2\mpl^2}\left({P+\rho\over H^2}-{\alpha_2\mpl^3\epsilon_{\dot\phi}\over \alpha_3 H}\right)
\,.
\ena
Since Eq. (\ref{eq_Z}) is the Riccati equation for $Z$, 
once we can find a special solution $Z_*(N)$,
we obtain a general solution as follows:\\
Setting $Z=Z_*+Y$, we find the Bernoulli equation 
as
\beann
{dY\over dN}+\left(Q_1-6Z_*)\right)Y=3Y^2
\,,
\enann
which can be linearized by setting $Y=1/X$ as
\beann
{dX\over dN}-\left(Q_1-6Z_*\right)X=-3
\,.
\enann
First we solve the homogeneous solution $X_H$, which satisfies
\beann
{dX_H\over dN}-\left(Q_1-6Z_*\right)X_H=0\,.
\enann
Using this homogenous solution, we obtain a 
general solution as 
\bea
X(N)=-3 X_H(N)\left[\int dN'{1\over X_H(N')}
\right]
\label{integral_X}
\,,
\ena
where
\bea
X_H(N)=\exp\left[\int^N dN' \left(Q_1(N')-6Z_*(N')\right))\right]
\,.
\label{integral_XH}
\ena
As a result, we obtain general solution for $Z$ as
\bea
Z(N)=Z_*(N)+{1\over X(N)}
\label{sol_Z}
\,.
\ena
Integrating Eq. (\ref{sol_Z}), we find the scalar field in terms of $N$ as
\bea
&&
\phi=\phi_0+{\mpl\over \alpha_3}\int_0^N dN' Z(N')
\label{sol_phi}
\,,
\ena
$\phi_0$ is the present value of the scalar field.

Solving the inverse problem given by Eq.  (\ref{sol_phi}), we find the $e$-folding $N$ in terms of $\phi$, i.e., $N=N(\phi)$.
As a result, inserting it in Eq. (\ref{basic_eq2}), we obtain the potential as
\beann
V(\phi)=-\rho(N(\phi))+3\mpl^2 H^2(N(\phi)) Z(N(\phi))^2
\,.
\enann

\subsection{Potential for $\Lambda$CDM model}
\label{potential_LCDM}
Now assuming matter dominant stage ($\rho=\rho_m$), we shall show 
the potential form for $\Lambda$CDM model,
which is given by
\beann
H^2={1\over 3M_{\rm F}^2}\left(\rho_m+\rho_{\rm vac} \right)
\,,
\enann
where $M_{\rm F}$ and $\rho_{\rm vac}$ are positive constants representing 
the modified Planck mass and the vacuum energy density, respectively.

To perform the integrations, we change the variable $N$ to $\xi$, which is defined 
by
\beann
\xi\equiv \sqrt{1+{\rho_m\over \rho_{\rm vac}}}\,.
\enann
Since the energy density is given by
\beann
\rho_m=\rho_{m,0}e^{-3N}
\,,
\enann
we find
\bea
d\xi=-{3(\xi^2-1)\over 2\xi}dN
\label{relation_xi_N}
\,.
\ena
Using Eq. (\ref{relation_xi_N}) and 
\beann
H&=&{\sqrt{\rho_{\rm vac}}\over \sqrt{3}M_{\rm F}}\xi\,,
\enann
we also find
\beann
Q_1&=&{3\over 2\xi^2}\left(\xi^2+2p\xi+1\right) 
\\
Q_2&=&-{3M_{\rm F}^2\over 2\mpl^2}{\xi^2-1\over \xi^2}+{3p\over \xi}
\,,
\enann
where
\beann
p\equiv {1\over 2\sqrt{3}}{\alpha_2\mpl M_{\rm F}\epsilon_{\dot \phi}
\over \alpha_3 \sqrt{\rho_{\rm vac}}}
\,.
\enann

The differential equation is 
\beann
&&
{dZ\over d\xi}-{(\xi^2+2p\xi+1)\over \xi(\xi^2-1)}Z+{2\xi\over \xi^2-1}Z^2
\\
&&~~~~~~~~~~
-{M_{\rm F}^2\over \mpl^2}{1\over \xi}+{2p\over \xi^2-1}=0
\,,
\enann
which is still the Riccati equation.

The equation for the scalar field and the potential are given by
\beann
&&
{d\phi\over d\xi}=-{2\mpl\over 3\alpha_3}{\xi\over \xi^2-1}\left(Z(\xi)-1\right) \,,
\enann
\beann
V&=&3\mpl^2H^2Z^2-\rho_m
\\
&=&V_0\left[\xi^2 Z^2(\xi)-{M_{\rm F}^2\over \mpl^2}\left(\xi^2-1\right)\right]
\,,
\enann
where 
\beann
V_0\equiv {\mpl^2\over M_{\rm F}^2}\,\rho_{\rm vac}
\,.
\enann

In order to find the analytic solution, we
 have to find a special solution $Z_*$.
It can be obtained by the hypergeometric functions.
However, since it is quite complicated, we may solve it numerically.

As for the initial condition, we shall consider
the limit of $\xi\rightarrow 1$ ($\rho_m\rightarrow 0$).
In this limit, 
$\Lambda$CDM model gives de Sitter expanding universe with 
$H =$ constant. 
If the potential $V$ is finite,  $Z$ is also finite.
As a result, $d\phi/dN$ must vanish in this limit.
It gives $Z\rightarrow 1$ as $\xi\rightarrow 1$.
In fact, we find the approximate solution by the power-series expansion near $\xi=1$ as
\beann
Z(\xi)&\approx& 1+z_1 (\xi-1)+z_2(\xi-1)^2 +\cdots\,,
\\
\phi(\xi)&\approx&\phi_1 (\xi-1)+\phi_2(\xi-1)^2 +\cdots\,,
\enann
where 
\beann
z_1&=&{1-r^2\over p-2}\,,~z_2={(r^2-1)[p^2-9p+2(r^2+6)]\over 2(p-3)(p-2)^2}\,,\cdots \,,\\
\phi_1&=&-{z_1\over 3}\frac{\mpl}{\alpha_3}\,,~\phi_2=-{z_1+2z_2\over 12}\frac{\mpl}{\alpha_3}\,,\cdots\,.
\enann
Here we define $r$ by 
\beann
r\equiv {M_{\rm F}\over \mpl}\,.
\enann
We then find the potential near $\phi=0$ as
\beann
V(\phi)=V_0\left[1-6(p-1)\phi+\cdots\right]\,.
\enann

We can also find the asymptotic solution in the limit of $\xi\rightarrow \infty$ as
\beann
Z(\xi)&\rightarrow& Z_\infty +c_Z \xi^{-\sqrt{1+8r^2}}+\cdots\,,\\
\phi(\xi)&\rightarrow&-{2\mpl\over 3\alpha_3}\left(Z_\infty -1\right)\ln \xi+\cdots\,,
\enann
where
\beann
Z_\infty\equiv {1+\sqrt{1+8r^2}\over 4}\,,
\enann
and $c_z$ is some constant.

Since the potential is given in this limit as
\beann
V\rightarrow V_0\left(Z_\infty^2-r^2\right)\xi^2
\,,
\enann
we find the asymptotic form of the potential as
\beann
V\approx V_0\left(Z_\infty^2-r^2\right)\exp\left[-{3\alpha_3\over (Z_\infty-1)\mpl}\phi\right]
\,,
\enann
which is the exponential potential (\ref{exp_pot}) with the exponent $\lambda$ given by
\beann
\lambda = -{3(\sqrt{1+8r^2}+3)\over 2(r^2-1)}
\,.
\enann

\begin{figure}[h]
	\includegraphics[width=7cm]{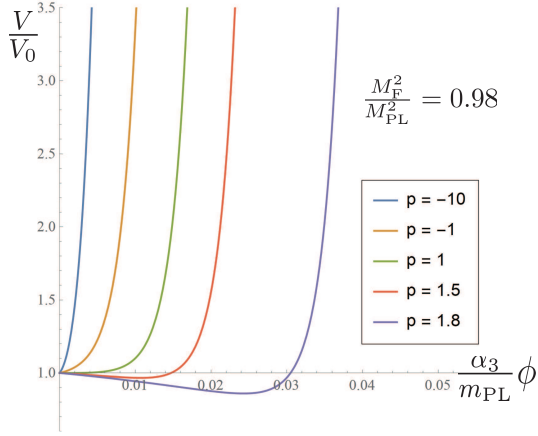}
	\caption{Potentials for $\Lambda$CDM model.
	We set $\Omega_{m,0}=0.3\,, \Omega_{\Lambda,0}=0.7$\,,
	and ${M_{\rm F}^2\over \mpl^2}=0.98$. 
		}
	\label{V_LCDM1}
\end{figure}

\begin{figure}[h]
		\includegraphics[width=7cm]{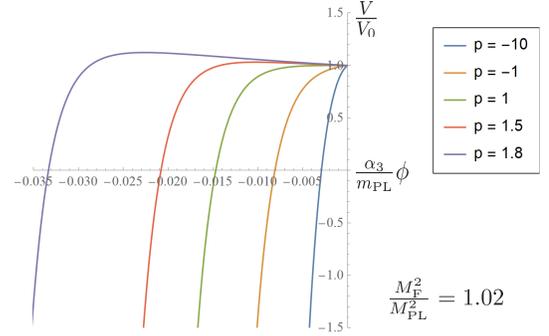}
	\caption{Potentials for $\Lambda$CDM model with  ${M_{\rm F}^2\over \mpl^2}=1.02$.
	The other parameters are the same as those in Fig. \ref{V_LCDM1}.  
	}
	\label{V_LCDM2}
\end{figure}

We show some numerical examples in Figs. \ref{V_LCDM1} and \ref{V_LCDM2}.
Here we assume that ${M_{\rm F}^2\over \mpl^2}=0.98$ or $1.02$, because 
the ``modified Planck'' mass $M_{\rm F}$ in the Friedmann equation should be close to the Planck mass 
$\mpl$.

Since $Z_\infty^2-r^2>0$ and $Z_\infty -1<0$ for $r<1$, while  $Z_\infty^2-r^2<0$ and $Z_\infty -1>0$ for $r>1$, we understand the above potential form with 
the fact that $dV/d\phi=-6(p-1)V_0$ at $\phi=0$.

 We cannot construct  numerically any appropriate potential for the parameter $p\geq 2$.

\begin{widetext}
\section{Density Perturbations and Effective Gravitational Constant}
\subsection{Basic equations for density perturbations}

According to Refs. \cite{DeFelice:2010as,DeFelice:2011hq,Tsujikawa:2014mba} we consider the perturbed metric on the flat FLRW background as
\bea
ds^2 = - (1 + 2 \Psi) dt^2 + 2 \partial_i \psi dx^i dt + a(t)^2 (1 + 2 \Phi) \delta_{ij} dx^i dx^j \,,
\ena
when $\psi = 0$, it corresponds to the Newtonian gauge. The energy-momentum tensor with perturbations are defined as
\bea
T^0_{~0} = - (\rho_m + \delta \rho_m) \,, \quad T^0_{~i} = - \rho_m \partial_i v_m \,, \quad T^i_{~j} = 0 \,,
\ena
where $v_m$ is a velocity potential of the perfect fluid. Note that we are considering only perturbations of nonrelativistic matter. 

Expanding the following action up to second order
\bea
S &=& \int d^4 x \sqrt{-g} \Big[ \frac{1}{2} \mpl^2 R + \alpha_2 \mpl^2 \sqrt{-X} + \alpha_3\mpl \ln \Big(-\frac{X}{\Lambda^4}\Big) \square \phi  - V(\phi)   + 3 \alpha_3^2 X \Big] + S_M (g_{\mu\nu}, \psi_M) \,.
\ena
Varying with respect to $\Psi$, $\Phi$, $\psi$, and $\delta\phi$, we find a set of equations in Fourier space as follows
\bea
E_{\Psi} &:& A_1 \dot\Phi + A_2 \dot{\delta\phi} + A_3 \frac{k^2}{a^2} \Phi + A_4 \Psi + A_5 \frac{k^2}{a^2} \psi + \Big(A_6 \frac{k^2}{a^2} - \mu\Big) \delta\phi - \delta\rho_m = 0 \,, \\
E_{\Phi} &:& B_1 \ddot \Phi + B_2 \ddot{\delta\phi} + B_3 \dot\Phi + B_4 \dot{\delta\phi} + B_5 \dot\Psi + B_6 \frac{k^2}{a^2} \Phi + 3 \nu \delta\phi +  \Big(B_8 \frac{k^2}{a^2} + B_9 \Big) \Psi \nn 
& &+ B_{10} \frac{k^2}{a^2} \dot\psi + B_{11} \frac{k^2}{a^2} \psi = 0 \,, \\
E_{\psi} &:& C_1 \dot\Phi + C_2 \dot{\delta\phi} + C_3 \Psi + C_4 \delta\phi + \rho_m v_m = 0 \,, \\
E_{\delta\phi} &:& D_1 \ddot \Phi + D_2 \ddot{\delta\phi} + D_3 \dot\Phi + D_4 \dot{\delta\phi} + D_5 \dot\Psi + D_6 \frac{k^2}{a^2} \dot\psi + D_8 \Phi + \Big( D_9 \frac{k^2}{a^2} - M^2 \Big) \delta\phi \nn 
& & + \Big( D_{10} \frac{k^2}{a^2} + D_{11} \Big) \Psi + D_{12} \frac{k^2}{a^2} \psi = 0 \,.
\ena
Components of the set of equations are
\beann
& &A_1 = 6 \mpl^2 H + 6 \mpl \alpha_3 \dot\phi \,, \quad A_2 = 6 \alpha_3 \mpl H+6 \alpha_3^2 \dot\phi \,, \quad A_3 = 2\mpl^2 \,, \\
& &A_4 = -6 \mpl^2 H^2-12 \alpha_3 \mpl H \dot\phi-\rho_m-6 \alpha_3^2 \dot\phi^2 \,, \quad A_5 = 2 \mpl^2 H+2 \alpha_3 \mpl \dot\phi \,, \\ 
& &A_6 = 2 \alpha_3 \mpl  \,, \quad \mu = V_{,\phi} \,, \\
& &B_1 = 6 \mpl^2 \,, \quad B_2 = 6 \alpha_3 \mpl \,, \quad B_3 = 18 \mpl^2 H \,, \\
& &B_4 = \frac{3 \alpha_2 \mpl^2 |\dot\phi|}{\dot\phi}-18 \alpha_3^2 \dot\phi \,, \quad B_5 = -6 \mpl^2 H-6 \alpha_3 \mpl \dot\phi \,, \quad B_6 = 2 \mpl^2 \,, \\
& &B_8 = 2 \mpl^2 \,, \quad B_9 = -6 \mpl^2 \dot H-18 \mpl^2 H^2-18 \alpha_3 \mpl H \dot\phi-6 \alpha_3 \mpl \ddot\phi+3 \rho_m \,, \\
& &B_{10} = 2 \mpl^2 \,, \quad B_{11} = 2 \mpl^2 H \,, \quad \nu = -V_{,\phi} \,, \\
& &C_1 = 2 \mpl^2 \,, \quad C_2 = 2 \alpha_3 \mpl \,, \quad C_3 = -2 \mpl^2 H-2 \alpha_3 \mpl \dot\phi \,, \quad C_4 = -6 \alpha_3 \mpl H+\frac{\alpha_2 \mpl^2 |\dot\phi|}{\dot\phi}-6 \alpha_3^2 \dot\phi \,, \\
& &D_1 = 6 \alpha_3 \mpl \,, \quad D_2 = 6 \alpha_3^2 \,, \quad D_3 = 36 \alpha_3 \mpl H-\frac{3 \alpha_2 \mpl^2 |\dot\phi|}{\dot\phi}+18 \alpha_3^2 \dot\phi \,, \\
& &D_4 = 18 \alpha_3^2 H \,, \quad D_5 = -6 \alpha_3 \mpl H-6 \alpha_3^2 \dot\phi \,, \quad D_6 = 2 \alpha_3 \mpl \,, \\
& &D_8 = 18 \alpha_3 \mpl \dot H-\frac{9 \alpha_2 \mpl^2 H |\dot\phi|}{\dot\phi}+54 \alpha_3 \mpl H^2+54 \alpha_3^2 H \dot\phi+18 \alpha_3^2 \ddot\phi-3 V_{,\phi} \,, \\
& &D_9 = 6 \alpha_3^2+\frac{8 \alpha_3 \mpl H}{\dot\phi}-\frac{\alpha_2 \mpl^2}{|\dot\phi|} \,, \quad D_{10} = 2 \alpha_3 \mpl \,, \\
& &D_{11} = -6 \alpha_3 \mpl \dot H-18 \alpha_3 \mpl H^2-18 \alpha_3^2 H \dot\phi-6 \alpha_3^2 \ddot\phi-V_{,\phi} \,, \\ 
& &D_{12} = 8 \alpha_3 \mpl H-\frac{\alpha_2 \mpl^2 |\dot\phi|}{\dot\phi}+6 \alpha_3^2 \dot\phi \,, \quad M^2 = V_{,\phi\phi} \,.
\enann

\end{widetext}
Note that $B_7 = D_7 = 0$. Since the matter is conserved, the perturbed energy-momentum tensor is satisfied
\bea
 \delta \nabla_{\mu} T^{\mu}_{~\nu} = 0 \,.
\ena

From these perturbation equations, we can also confirm that this theory has two degrees of freedom.
Although the perturbation equations contain $\dot{\delta\phi} $ and $\ddot{\delta\phi} $  as well as 
$\delta \phi$, we can eliminate those derivative terms by combining the perturbation equations, and 
obtain $\delta \phi$ in terms of the perturbation variables of matter fluid and metric components
 ($\delta \rho_m, v, \Phi, \Psi$, and $\psi$) and those time derivatives.
 Hence the perturbation of the scalar field is algebraically determined by the other perturbation variables.
 There is no additional degree of freedom coming from the scalar field.

Choosing the Newtonian gauge, the components $\nu = 0$ and $\nu = i$ lead to
\bea
\dot{\delta\rho_m} + 3H \delta\rho_m + \frac{k^2}{a^2} \rho_m v_m + 3 \rho_m \dot\Phi &=& 0 \,,\label{nu0} \\
\dot v_m &=& \Psi \,, \label{nui}
\ena
respectively. The useful combination is 
\bea
3(\dot E_{\psi} + 3 H E_{\psi}) - E_{\Phi} = 0 \,,
\ena
with the basic equations of the flat FLRW background and Eq. (\ref{nui}), the above relation becomes
\bea
B_6 \Phi + B_8 \Psi = 0 \,. \label{Egamma}
\ena 
We are interested in the subhorizon regime, $k^2 /a^2 \gg H^2$, and using the quasistatic approximation, i.e. the dominant contributions terms are $k^2/a^2$, $\delta \rho_m$, and $M^2$. We also neglect the oscillating term of $\delta \phi$ and assume that the variations on gravitational potentials are small. Thereby, the $E_{\Psi}$ and the $E_{\delta \phi}$ become
\bea
& &A_3 \frac{k^2}{a^2} \Phi + A_6 \frac{k^2}{a^2} \delta\phi - \delta\rho_m \simeq 0 \,, \label{EPsi} \\
& &\Big( D_9 \frac{k^2}{a^2} - M^2 \Big) \delta\phi + D_{10} \frac{k^2}{a^2}\Psi \simeq 0 \,. \label{EPhi}
\ena
Solving Eqs. (\ref{Egamma}), (\ref{EPsi}), and (\ref{EPhi}) we find
\bea
\frac{k^2}{a^2} \Psi &\simeq& - \frac{\Big(B_6 D_9 \frac{k^2}{a^2} - B_6 M^2\Big) \delta \rho_m}{\Big(A_6^2 B_6 + B_8^2 D_9\Big)\frac{k^2}{a^2} - B_8^2 M^2} \,, \label{EGeff}\\ 
\Phi &=& - \Psi \,.
\ena
Under the above approximations, taking time derivative on Eq. (\ref{nu0}) and using Eq. (\ref{nui}) and the conservation of matter density equation, equation of motion of the density contrast is given by
\bea
\ddot \delta_m + 2 H \dot\delta_m + \frac{k^2}{a^2} \Psi = 0 \,,
\ena 
where the density contrast is defined as $\delta_m = \delta\rho_m / \rho_m$.

Substituting Eq. (\ref{EGeff}) into above equation we find 
\bea
\ddot \delta_m + 2 H \dot\delta_m - 4 \pi G_{\rm eff} \rho_m \delta_m \simeq 0 \,,
\ena
where the effective gravitational constant is
\beann
G_{\rm eff} &=& \frac{2\mpl^2 \Big(B_6 D_9 \frac{k^2}{a^2} - B_6 M^2\Big)}{\Big(A_6^2 B_6 + B_8^2 D_9\Big)\frac{k^2}{a^2} - B_8^2 M^2}  G_{\rm N}
\nn
&=& \frac{\Big(6\alpha_3^2 + \frac{8\alpha_3\mpl H}{\dot\phi} - \frac{\alpha_2 \mpl^2}{|\dot\phi|} - M^2 \frac{a^2}{k^2}\Big)}{\Big(8\alpha_3^2 + \frac{8\alpha_3\mpl H}{\dot\phi} - \frac{\alpha_2 \mpl^2}{|\dot\phi|} - M^2\frac{a^2}{k^2}\Big)} G_{\rm N}\,.
\enann
Here we set $G_{\rm N}\equiv {1\over 8\pi \mpl^2}$.

\begin{widetext}

In the subhorizon limit the term $M^2 a^2 / k^2$ is very small compared to other terms, then we can neglect this term. Using the new Hubble parameter and Eq. (\ref{newHsgn}) the effective gravitational constant becomes
\bea
G_{\rm eff} &=&
 \left\{1 - \frac{2{\alpha_3\over \mpl}\frac{d \phi}{dN} (\rho_m + 2V - \frac{\mpl}{3\alpha_3} V_{,\phi})}{\Big(1+ \frac{\alpha_3}{\mpl}\frac{d\phi}{dN}\Big)\Big[8\Big(\rho_m + 2V - \frac{\mpl}{3\alpha_3} V_{,\phi}\Big) - {\alpha_2^2\over \alpha_3^2} \mpl^4\Big]}\right\} G_{\rm N}\,,
\ena
where we neglect contributions from radiation. 

\end{widetext}
The gravitational slip parameter is defined as
\bea
\eta \equiv -\frac{\Phi}{\Psi} \,,
\ena
which is always equal to one in this model.

\subsection{Effective gravitational constant and observational constraints}

Since the effective gravitational ``constant'' is time-dependent, we have to
take into account  the observational constraints.
The lunar-laser ranging experiment \cite{Hofmann:2018myc,Biskupek:2020fem} gives
the constraint such that 
\bea
{\dot G\over G}&=&(-5.0\pm 9.6)\times 10^{-15}~{\rm yr}^{-1} \,,
\label{dot_G}
\\
{\ddot G\over G}&=&(1.6\pm 2.0)\times 10^{-16}~{\rm yr}^{-2}
\,.
\label{ddot_G}
\ena

If the gravitational constant evolves due to the cosmic expansion, we expect that 
$\dot G_{\rm N}/G_{\rm N}\sim O(H_0)\sim 7\times 10^{-11}\, {\rm yr}^{-1}$
 and  $\ddot G_{\rm N}/G_{\rm N}\sim O(H_0^2)\sim 5\times 10^{-21}\, {\rm yr}^{-1}$.
As a result,  the condition (\ref{dot_G}) will give a strong constraint,
 but  the constraint (\ref{ddot_G}) may be much weaker.

\subsubsection{Exponential potential with $\lambda>6$}

Assuming the exponential potential with $\lambda>6$ discussed in \S. \ref{exp_pot_lam>6},
we show the behaviour of $G_{\rm eff}(z)$.
We consider only $+$ branch solution with $\epsilon_V=1$.
We then find the effective gravitational ``constant'' is given by
\beann
{G_{\rm eff}\over G_{\rm N}}=1-{2(Z_+-1)H_+\epsilon_{\dot\phi}
\over 8H_+ Z_+\epsilon_{\dot\phi}-{\alpha_2\over \alpha_3}\mpl}
\,,
\enann
where
\beann
Z_+&=&{F_+\over S_+\sqrt{D}} 
\\
H_+^2&=&{V_\infty\over 6\mpl^2}{DS_+^2R_+\over F_+^2} \,.
\enann

\begin{figure}[h]
	\includegraphics[width=7cm]{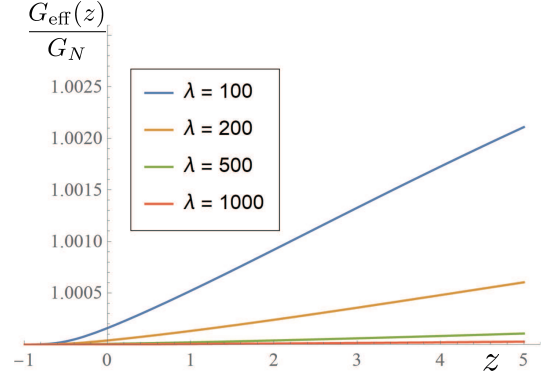}
	\caption{Evolutions of $G_{\rm eff}/G_{\rm N}$  for the cases of $\lambda=100, 200,  500$ and 1000 in terms of the redshift $z$.
	We choose the same parameter values as those in Fig. \ref{H}. }
	\label{Geff}
\end{figure}

\begin{figure}[h]
	\includegraphics[width=7cm]{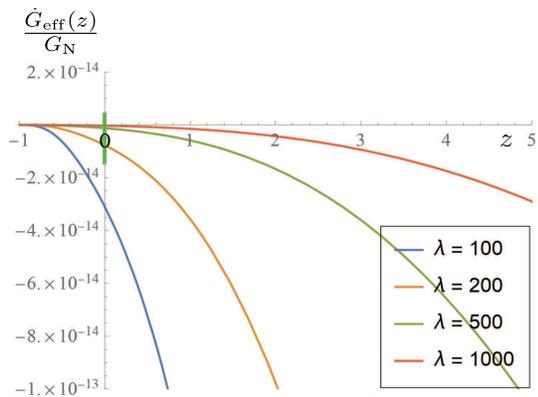}
	\caption{Evolutions of $\dot G_{\rm eff}$ in terms of the redshift $z$.
	We choose the same parameter values as those in Fig. \ref{Geff}.
		The constraint from lunar-laser ranging experiment, Eq. (\ref{dot_G}), is  
		given by the green line segment at $z=0$. }
	\label{dotGeff}
\end{figure}

In Fig. \ref{Geff}, we depict  the evolution of $G_{\rm eff}/G_{\rm N}$.
Taking the time derivative of $G_{\rm eff}$, we show 
the behaviour of 
$\dot G_{\rm eff}/G_{\rm N}$ in terms of the redshift $z$
in Fig. \ref{dotGeff}.

In order to satisfy the constraint (\ref{dot_G}), we
 find $ \lambda \geq 145$, which corresponds to 
\beann
0> {\dot G_{\rm eff}\over G_{\rm N}}\Big{|}_0\geq -1.458\times 10^{-14}
\,,
\enann
where  $\dot G_{\rm eff}/G_{\rm N}|_0$ is the present value.
Hence the constraint obtained from the big-bang nucleosynthesis
($\lambda\geq 208$) gives the sufficient condition.

The constraint (\ref{ddot_G}) on $\ddot G/G$  is always satisfied for any values of $\lambda$ as we expected.

\subsubsection{The potential for $\Lambda$CDM background universe}

If the potential is given by one discussed in \S. \ref{potential_LCDM}, 
we recover $\Lambda$CDM model for 
the background dynamics.
However the effective gravitational ``constant'' is no longer constant.
It depends on time as
\beann
{G_{\rm eff}\over G_{\rm N}}=1-{2(Z-1)H\epsilon_{\dot\phi}
\over 8H Z\epsilon_{\dot\phi}-{\alpha_2\over \alpha_3}\mpl}
\,.
\enann

\begin{figure}[h]
	\includegraphics[width=5cm]{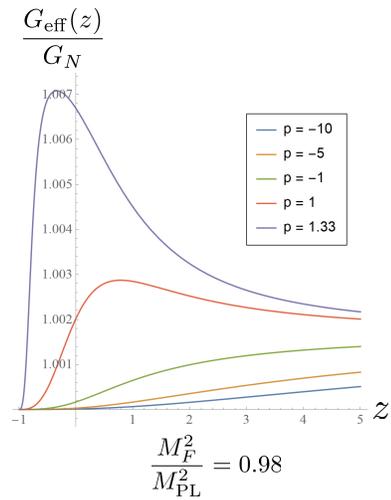}\\[2em]
	\includegraphics[width=5cm]{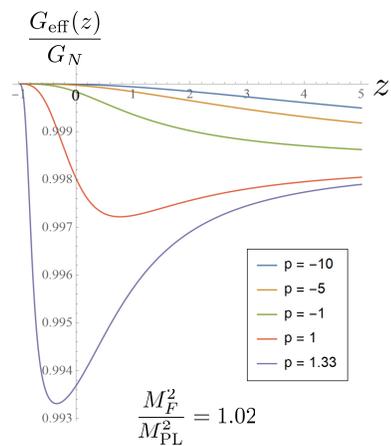}
	\caption{Evolutions of $G_{\rm eff}$  for the cases of $p=1.33, 1 -1, -5$ and $-10$  in terms of the redshift $z$. The top and bottom figures correspond to  ${M_{\rm F}^2\over \mpl^2}=0.98$  and 
	 ${M_{\rm F}^2\over \mpl^2}=1.02$, respectively.}
	\label{Geff_LCDM}
\end{figure}

We show the evolution of $G_{\rm eff}$ in terms of the redshift $z$ in Fig. \ref{Geff_LCDM}.
We consider only the cases of $p\leq 1.33$ because  $G_{\rm eff}$ will diverges at some value of $z$ when $p\gsim 1.34$.

\begin{figure}[h]
	\includegraphics[width=6cm]{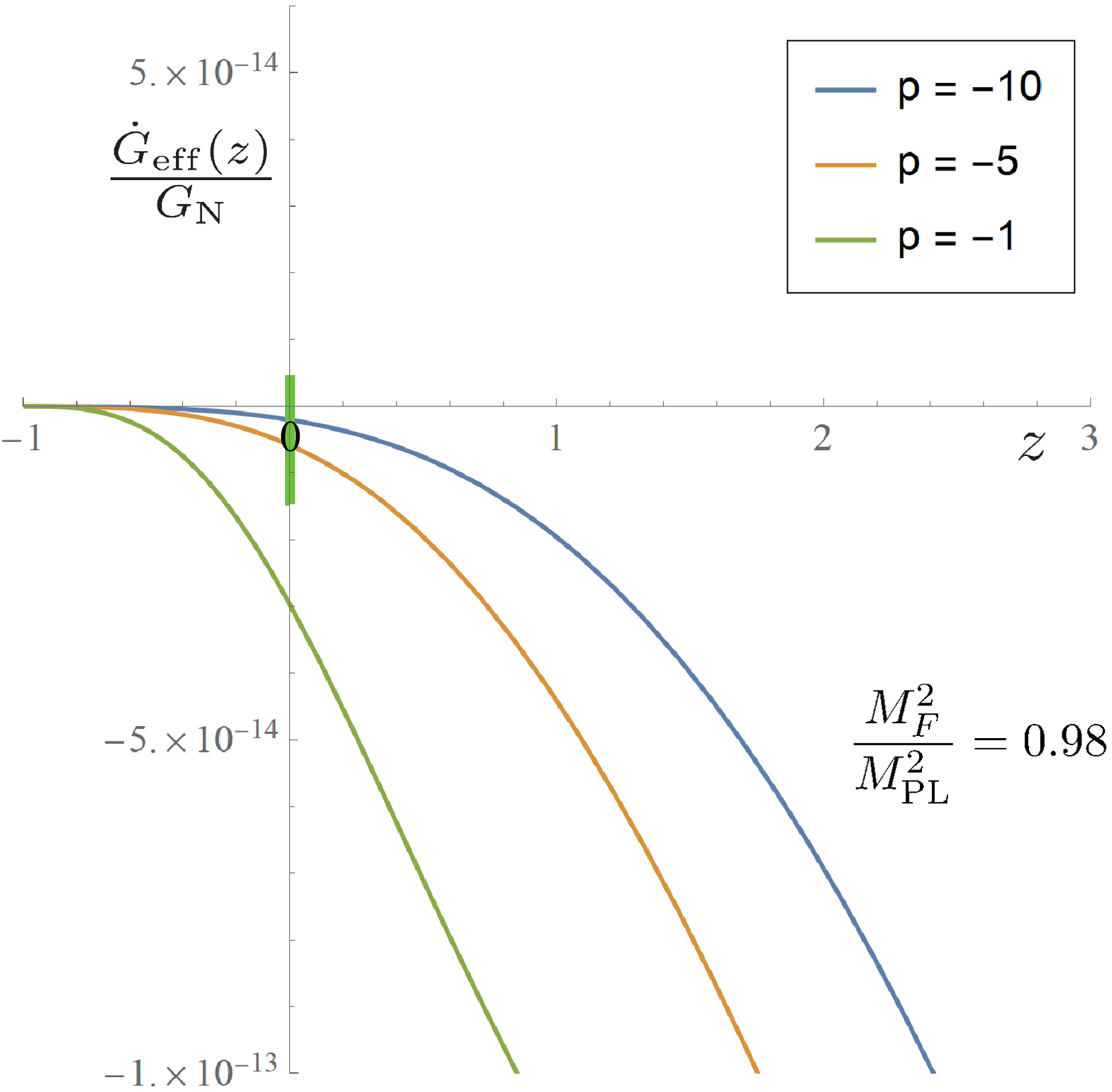}\\[2em]
	\includegraphics[width=6cm]{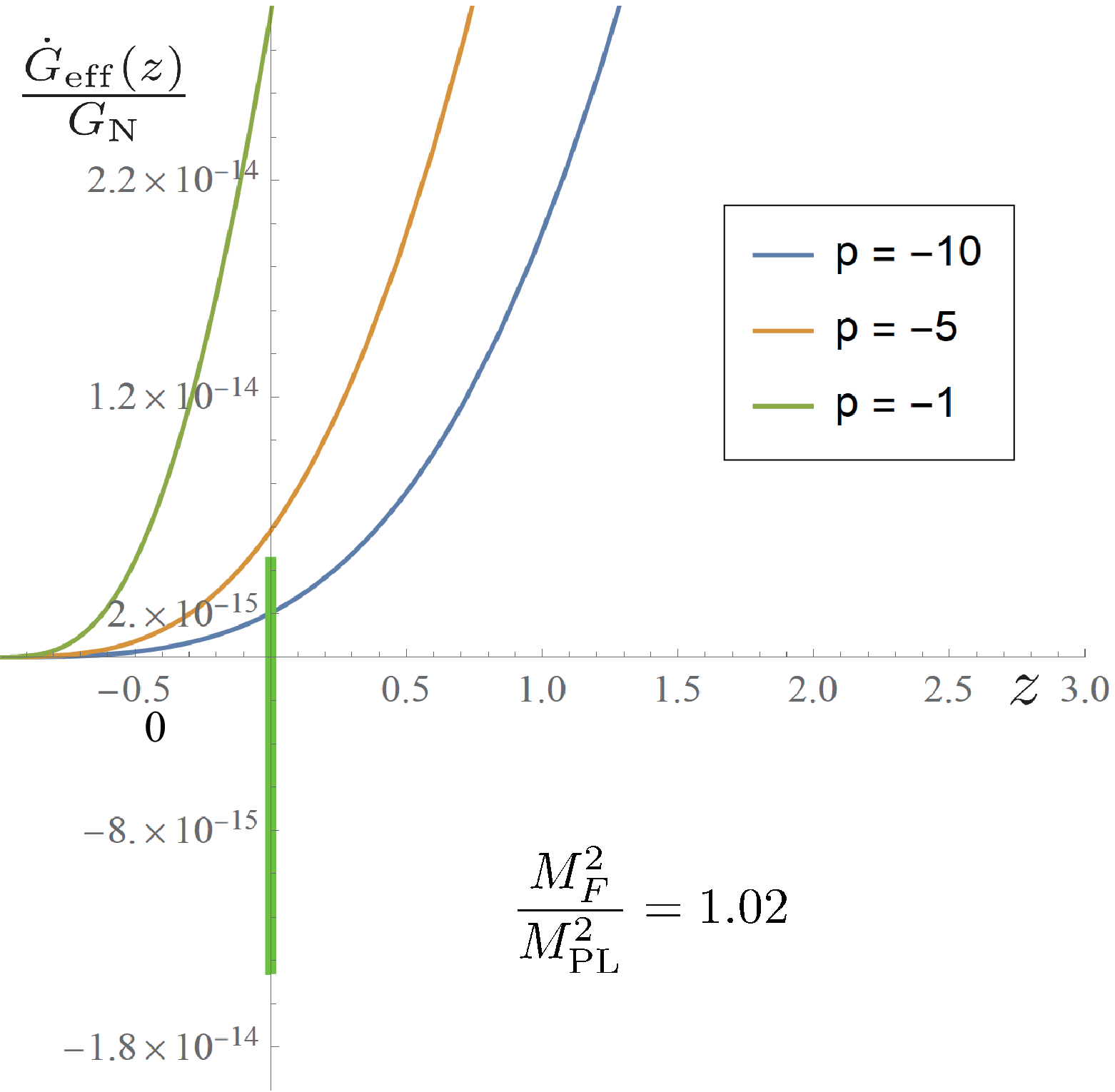}
	\caption{Evolutions of $\dot G_{\rm eff}$ for the cases of $p=-1, -5$ and $-10$ in terms of the redshift $z$.
	The constraint from lunar-laser ranging experiment, Eq. (\ref{dot_G}), is  
		given by the green line segment at $z=0$.
		The top and bottom figures correspond to  ${M_{\rm F}^2\over \mpl^2}=0.98$  and 
	 ${M_{\rm F}^2\over \mpl^2}=1.02$, respectively.}
	\label{dotGeff_LCDM}
\end{figure}

We can also discuss the time evolution of $\dot G_{\rm eff}$, 
which plots are given in Fig. \ref{dotGeff_LCDM}.

These figures show that when we decrease the value of $p$, 
the present value of $\dot G_{\rm eff}$ becomes smaller.
From the observational constraint (\ref{dot_G}), we find 
\bea
p \leq 
\left\{
\begin{array}{ccc}
-2.4~& {\rm for} & {M_{\rm F}^2/ \mpl^2}=0.98
\\[.5em]
-6.0  ~& {\rm for} &~ {M_{\rm F}^2/ \mpl^2}=1.02
 \,.
 \\
\end{array}
\right.
\label{constraint_a2a3}
\ena

The constraint (\ref{ddot_G}) is automatically satisfied for any values of $p$.

The above constraints (\ref{constraint_a2a3}) correspond to 
\beann
{\alpha_2\over \alpha_3 } > 8.4 \sqrt{{\rho_{\rm vac}\over  \mpl^4}} 
\enann
for $ {M_{\rm F}^2/ \mpl^2}=0.98$, and
\beann
{\alpha_2\over \alpha_3} > 20.6 \sqrt{{\rho_{\rm vac}\over \mpl^4}}
\enann
for $ {M_{\rm F}^2/ \mpl^2}=1.02$.

\section{Discussion and Remarks}
\label{conclusion}

We discuss  a Cuscuta-Galileon gravity theory, 
which is one simple extension of a Cuscuton gravity theory and 
still preserves two degrees of freedom.
We apply it to cosmological model and 
present the effective Friedmann equation assuming the  flat FLRW metric.
Although there exists no additional degrees of freedom, 
introduction of a potential of a scalar field changes the dynamics.
The scalar field is completely determined by matter fields.

Giving an exponential potential as an example, 
we discuss the evolution of the Hubble expansion  parameter. 
Since the gravitational ``constant'' $G_{\rm F}$ in the effective Friedmann equation
 becomes  time-dependent.
we restrict the parameters in our models  with the constraint by the big-bang nucleosynthesis.

We also present how to construct a potential once we know the
 evolution of the  Hubble parameter. 
As an example, we present  the potential form
 to obtain the $\Lambda$CDM cosmology for the background evolution.

We then analyze the density perturbations, which equation is  characterized 
only by a change of the gravitational ``constant'' $G_{\rm eff}$. 
Note that $G_{\rm eff}$ in the above $\Lambda$CDM model is also time-dependent.
Hence it is not exactly the same as the $\Lambda$CDM cosmology in GR.
We then  restrict the parameters in our models using  the observational constraints by the lunar-laser-ranging experiment.

In the case of exponential potential, there appears
the time-dependence of the gravitational constant in the effective Friedmann equation, 
which may give a chance to explain the Hubble tension problem
\cite{Planck:2013pxb, Planck:2018vyg, Riess:2019cxk, Riess:2020fzl, DiValentino:2021izs}.

\begin{figure}[h]
		\includegraphics[width=7cm]{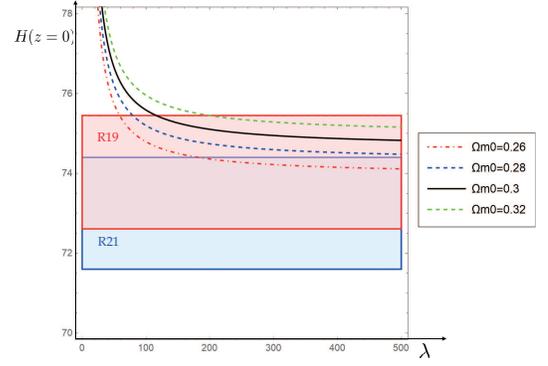} 
	\caption{The present value of the Hubble expansion rate in terms of $\lambda$.
	The dashed green line, black solid line, dashed blue line, and dot-dashed red line 
	correspond to $\Omega_{\rm m,0}=0.32, 0.3, 0.28$ and $0.26$, respectively.
	Two observational data by R19 \cite{Riess:2019cxk} and by R21 \cite{Riess:2020fzl} are given by the red shaded and blue shaded regions, respectively.
	 }
	\label{H(z=0)}
\end{figure}

As shown in
Fig. \ref{HvsLCDM},
 the Hubble expansion rate at  $z\leq 1$ is about 10\% larger 
than the value of  the $\Lambda$CDM model.
 We then plot the present value of the Hubble expansion rate in terms of $\lambda$  in Fig. \ref{H(z=0)}.
For the reference, 
 we also show the observational data R19 of the Hubble expansion rate near $z=0$, which is 
 obtained  from observations of 70 long-period Cepheids in the Large Magellanic Cloud\cite{Riess:2019cxk}.
This figure shows that our model with  $\Omega_{\rm m,0}=0.3$  is consistent with the observational data R19 if $\lambda>117$,
which should be satisfied from the constraint by nucleosynthesis ($\lambda>208$).
If we take the observational data R21,  which  is determined from observations of 75 Milky Way Cepheids
\cite{Riess:2020fzl},  it strays from the allowed range.
However the result depends on the density parameter $\Omega_{\rm m,0}$.
If $\Omega_{\rm m,0}\lsim 0.28$, our model with large $\lambda$ is still consistent with R21 as well as R19.
 
\begin{figure}[h]
		\includegraphics[width=7cm]{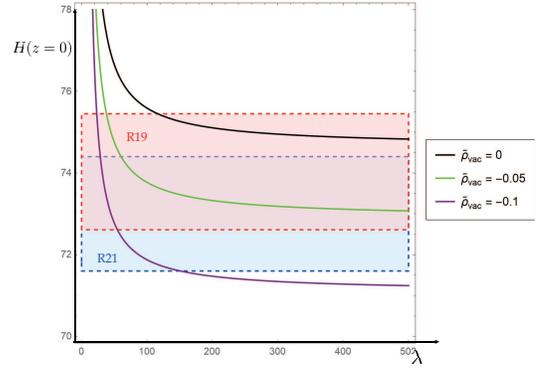} 
	\caption{The present value of the Hubble expansion rate in terms of $\lambda$
	when we add a vacuum energy $\rho_{\rm vac}$.
	We assume $\Omega_{\rm m,0}=0.3$.
	The black line, green line,  and purple line 
	correspond to $\tilde \rho_{\rm vac}=0, -0.05 $ and $-0.1$, respectively, where 
	$\tilde\rho_{\rm vac} = \rho_{\rm vac}/V_{\infty}$.
	Two observational data by R19 \cite{Riess:2019cxk} and by R21 \cite{Riess:2020fzl} are given by the red shaded and blue shaded regions, respectively.
	 }
	\label{H(z=0)_V0}
\end{figure}

Our model will be improved when we add a negative vacuum energy $\rho_{\rm vac}$ as well as matter and radiation densities, $\rho_m$ and $\rho_r$.
The effective Friedmann equation is given in Appendix \ref{expo_potential_cosconstant}.
Assuming $\Omega_{\rm m,0}=0.3$, we plot the present value of the Hubble expansion rate in terms of $\lambda$  in Fig. \ref{H(z=0)_V0}.
The case with $\rho_{\rm vac}=-0.05V_\infty $ fits well both for R19 and R21, 
where $V_\infty={3\alpha_2^2 \over (\lambda-6)^2\alpha_3^2}\mpl^4$.
Such a small negative vacuum energy might be obtained in the context of string theory \cite{Demirtas:2021ote}.

When we take the limit of 
$\lambda\rightarrow \infty$ and $\alpha_3\rightarrow 0$
with keeping $V_\infty$  finite, 
we obtain the same results as those in the original Cuscuton theory with an exponential potential, which Friedmann equation is given by Eq. (\ref{Friedmann_org_exp}). 
Since our model could be successful to explain the history of our universe when $\lambda$ is large, the original 
Cuscuton theory with an exponential potential may also have the possibility to 
solve the Hubble tension 
problem. In fact, the present Hubble constant becomes  $H_0=74.65$ km/s/Mpc  when we normalize 
the Hubble   parameter at $z=1100$ by use of 
the CMB data based on the $\Lambda$CDM universe model.
This is quite close to the value
in our model with large $\lambda$.
One difference is that two ``gravitational constants``, $G_{\rm F}$ and $G_{\rm eff}$,   
are exactly the same as $G_{\rm N}$ in the original Cuscuton theory.

In the case of the potential for the $\Lambda$CDM universe discussed in \S.\ref{potential_LCDM}, we  also find the cosmological model in the Cuscuton theory as the limiting case of our Cuscuta-Galileon theory. In fact, if we take the limit of $p\rightarrow -\infty$ as well as
$\alpha_3\rightarrow 0$ keeping $p\alpha_3$ finite, 
the constructed potential in \S.\ref{potential_LCDM} 
becomes a quadratic function of the scalar field $\phi$
(see Appendix \ref{quadratic_pot}).

The above two examples suggest that our cosmological model
includes that in the original Cuscuton theory as the limiting case.
The difference is $G_{\rm eff}$, which is time-dependent in our 
model, while that in the original Cuscuton theory is constant ($G_{\rm N}$).

Although we may explain the present large Hubble constant by the observation of  
nearby SNe Ia as well as small value obtained from CMB data assuming $\Lambda$CDM model,
we may have to analyze our model more carefully from the observational view points.
Even if it turns out that the present model with the exponential potential is not consistent with 
observational data, we still have many possibilities.
We may find a better model by tuning the potential as shown in the construction method (\S. \ref{construction}).
We can also extend our Cuscuta-Galileon gravity theory \cite{Iyonaga:2018vnu,Iyonaga:2020bmm} because our model is the simplest one.
We may obtain a better theory for observations.
We shall leave these analyses as future works.

\section*{Acknowledgments}

K.M. would like to thank Antonio De Felice, Shinji Mukohyama, and Masroor C. Pookkillath  
for useful comments and fruitful discussions.
K.M. also acknowledges 
 the Yukawa Institute for Theoretical Physics at
Kyoto University, where  most of the present work was completed during the
Visitors Program of FY2021.
This work was supported in part 
by JSPS KAKENHI Grants No. JP17H06359 and No. JP19K03857
and 
by a Waseda University Grant for Special Research Project (No. 2021C-569)

\begin{widetext}

\newpage

\appendix

\section{rescaling of scalar field}
\label{rescaling}
 In the present Cuscuta-Galileon model defined by the action (\ref{action0}), 
without loss of generality, unless $\alpha_3=0$, 
we can always set $\alpha_3=1$ by rescaling the scalar field $\phi$ as $\tilde\phi=\alpha_3\phi$.
In fact, defining 
\beann
\tilde X\equiv  g^{\mu\nu}\partial_{\mu}\tilde \phi \partial_{\mu} \tilde \phi =\alpha_3^2 X
\,,
\enann
we find that 
 the above action $S$ is given by
\beann
S &=& \int d^4 x \sqrt{-g} \Big[ \frac{1}{2} \mpl^2 R +{\alpha_2\over \alpha_3} \mpl^2\sqrt{-\tilde X} +  \ln \Big(-\frac{\tilde X}{\alpha_3^2\Lambda^4}\Big) \square\tilde  \phi  - V(\tilde \phi/\alpha_3)   +3\tilde X \Big] + S_M (g_{\mu\nu}, \psi_M) \,, \nn
\enann
Introducing the scaled parameters as
\beann
\tilde \alpha_2={ \alpha_2\over \alpha_3}\,,~\tilde \Lambda ^4= \alpha_3^2\Lambda^4 \,,
\enann
and redefining the potential as
\beann
\tilde V(\tilde \phi)=V(\tilde \phi/\alpha_3)
\,,
\enann
we find
\bea
S &=& \int d^4 x \sqrt{-g} \Big[ \frac{1}{2} \mpl^2 R +\tilde \alpha_2\mpl^2 \sqrt{-\tilde X} +  \ln \Big(-\frac{\tilde X}{\tilde \Lambda^4}\Big) \square\tilde  \phi  -\tilde V(\tilde \phi)   +3\tilde X \Big] + S_M (g_{\mu\nu}, \psi_M) \,, \nn
\ena
which is the action (\ref{action0}) with $\alpha_3=1$.

\end{widetext}
\appendix
\setcounter{section}{1}
\section{Original Cuscuton Gravity ($\alpha_3=0$)}
Here we reanalyze cosmological dynamics in the original Cuscuton gravity with a
potential ($\alpha_3=0$).
The basic equations are given by
\bea
&& H {\rm sgn}(\dot \phi)=-{1\over 3\alpha_2 \mpl^2}V_{,\phi} \,,
\label{EOM_cuscuton1}
\\
&& H^2={1\over 3\mpl^2}\left(\rho+V\right)
\label{EOM_cuscuton2}
\,,
\ena
where $\rho=\rho_m+\rho_r$.

We then discuss two potentials, the quadratic potential and the exponential potential as
analyzed in \cite{Afshordi:2006ad} and \cite{Afshordi:2007yx}.

\subsection{Quadratic potential}
\label{quadratic_pot}
We first assume the potential is given by
\beann
V=V_0+\frac{1}{2}m^2\phi^2 \,.
\enann
In this case, since $V_{,\phi}=m^2\phi$, 
we have a constraint such that
\beann
{1\over 3\mpl^2}\left(\rho+V_0+{1\over m^2}\phi^2\right)={1\over 9\alpha_2^2\mpl^4}m^4\phi^2 \left(=H^2\right)
\,,
\enann
which
gives
\beann
\phi^2={6\alpha_2^2\mpl^2(\rho+V_0)\over m^2(2m^2-3\alpha_2^2\mpl^2)}
\,.
\enann
Using this relation, we find the Friedmann equation as
\bea
H^2={1\over 3M_{\rm F}^2}\left(\rho+V_0\right)
\label{cuscuton_LCDM}
\,,
\ena
where 
\bea
M_{\rm F}^2\equiv \left(1-{3\alpha_2^2\mpl^2\over 2m^2}\right)\mpl^2
\label{relation_MFMPL}
\,.
\ena
Eq. (\ref{cuscuton_LCDM}) describes the $\Lambda$CDM model with
new gravitational constant 
\bea
G_{\rm F}\equiv {G_{\rm N}\over \left(1-{3\alpha_2^2\mpl^2\over 2m^2}\right)} \, 
(>G_{\rm N})\,.
\ena
Since the gravitational constant in the Friedmann equation must be close to 
the Newtonian gravitational constant $G_{\rm N}$, we have a constraint
\beann
m^2\gg{3\alpha_2^2\over 2}\mpl^2 \,.
\enann
\subsection{Exponential potential}
Next we consider the exponential potential
\beann
V=\epsilon_V \mpl^4 \exp\left(\lambda\phi/\mpl\right)\,.
\enann
The constraint equation (\ref{constraint}) with $\alpha_3=0$ is
\beann
&&
{1\over 3}\left[\rho+\epsilon_V \mpl^4 \exp\left(\lambda\phi/\mpl\right) \right]
\\
&=&
{1\over 9\alpha_2^2\mpl^2}V_{,\phi}^2={\lambda^2\mpl^4 \over 9\alpha_2^2}\exp\left(2\lambda\phi/\mpl\right) 
\,.
\enann
By setting 
$
\chi\equiv \exp\left(\lambda\phi/\mpl\right),
$
we find the quadratic equation for $\chi$ as
\bea
\chi^2-{3\epsilon_V \alpha_2^2\over \lambda^2}\chi-{3\alpha_2^2\over \lambda^2\mpl^4}\rho=0\,.
\label{eq_xi}
\ena
In order to have real positive roots for this equation,
we find the condition such that
\beann
\left({3\epsilon_V \alpha_2^2\over \lambda^2}\right)^2+{12\alpha_2^2\over \lambda^2\mpl^4}\rho\geq 0\,,
\enann
which is always satisfied because
$
\rho \geq 0
$
\,.

The solution for Eq. (\ref{eq_xi}) is
\beann
\chi=\chi_+(\rho) \equiv {3 \alpha_2^2\over 2\lambda^2}\left(
\epsilon_V+ \sqrt{1+{4\lambda^2\over 3\alpha_2^2\mpl^4}\rho}\right)
\,.
\enann
Only a $+$ branch of solutions is possible because $\chi$ should be positive. 
Note that $\epsilon_V=\pm 1$. 

\begin{widetext}

We then find the scalar field $\phi$ in terms of $\rho$ as
\beann
\phi=\phi_+\equiv {\mpl\over \lambda}\ln\left[{3\alpha_2^2\over 2\lambda^2}\left(
\epsilon_V + \sqrt{1+{4\lambda^2\over 3\alpha_2^2\mpl^4}\rho}\right)
\right] \,.
\enann
As a result,  the Friedmann equation (\ref{EOM_cuscuton2}) is given by
\bea
H_+^2={1\over 3\mpl^2}\left[
\rho+{3 \alpha_2^2\mpl^4 \over 2\lambda^2}\left(
1+ \epsilon_V \sqrt{1+{4\lambda^2\over 3\alpha_2^2\mpl^4}\rho}\right)\right] \,.
\label{Friedmann_org_exp}
\ena
\end{widetext}

In the early stage ($\rho \rightarrow \infty$), 
the universe starts from  the radiation dominant stage 
and follows by the matter dominant stage both for $\epsilon_V=\pm 1$.

For the late stage, we discuss the cosmic evolution for 
two cases ($\epsilon_V = \pm 1$) separately.
\subsubsection{$\epsilon_V = + 1$ {\rm (positive potential)}}

In the limit of $\rho\rightarrow 0$, we obtain
\beann
3\mpl^2 H_+^2 = {3 \alpha_2^2 \over \lambda^2}\mpl^4\equiv \rho_{\rm DE} ~(>0)\,,
\enann
which gives de Sitter expansion with the Hubble expansion rate $H_{\rm DE} = |\alpha_2| \mpl / |\lambda|$.
For the present acceleration, we have to impose the condition such that
\bea
{|\alpha_2| \over |\lambda|}\sim O(10^{-60})\ll 1\,.
\ena

\subsubsection{$\epsilon_V = - 1$ {\rm (negative potential)}}

In this case, in the limit of $\rho\rightarrow 0$, 
we find the Friedmann equation as
\beann
3\mpl^2 H_+^2=\frac{\lambda^2 \rho^2}{3 \alpha_2^2 \mpl^4 } \,, ~~~ {\rm and } ~~~ \rho \sim \rho_m \,,
\enann
which gives
\beann
a(t) \propto t^{1\over 3}\,.
\enann
 This is the expansion law for the stiff matter ($P=\rho$) in GR.

Consequently,
only  the case of $\epsilon_V = +1$ (positive exponential potential) 
provides the big-bang universe followed by an accelerating expansion.

\subsection{Construction of appropriate potential}
\label{construction0}
We may construct an appropriate potential once we know the expansion of the universe from observation.
Here we provide how to construct the potential giving the Hubble expansion parameter 
$H$ in terms of the redshift $z$.

From basic equations 
we find
\bea
V_{,\phi}^2&=&9\alpha_2\mpl^4 H^2 \,,
\label{cuscuton_Vphi}
\\
V&=&3\mpl^2H^2-\rho
\,.\label{cuscuton_V}
\ena
We rewrite Eq. (\ref{cuscuton_Vphi}) in terms of $z$ as
\beann
\left({d\phi\over  dz}\right)^2=\left({{dV/ dz}\over V_{,\phi}}\right)^2={1\over 9\alpha_2^2\mpl^4H^2}\left({dV\over  dz}\right)^2
\,.
\enann
From Eq. (\ref{cuscuton_V}), we obtain
\beann
{dV\over dz}=6\mpl^2 H{dH\over dz}-{d\rho\over dz} \,,
\enann
then
\beann
{d\phi\over dz}&=&\pm {1\over 3|\alpha_2|\mpl^2 H(z)}{dV\over dz}  \\
&=&\pm{1\over |\alpha_2|}\left(2{dH\over dz}-{1\over 3\mpl^2 H}{d\rho\over dz}\right)\,.
\enann
Integrating this equation, we find $\phi=\phi(z)$. 
Solving $z=z(\phi)$ as the inverse problem, 
and inserting it into Eq. (\ref{cuscuton_V}), we find the potential 
$V(\phi)$.

In order to show it more explicitly, in what follows, we assume $\rho=\rho_m$.
Since 
\beann
{d\rho_m\over dz}={3\over 1+z}\rho_m
\,,
\enann
we find
\beann
{d\phi\over dz}
&=&\pm{1\over |\alpha_2|}\left(2{dH\over dz}-{\rho_m\over \mpl^2 (1+z) H}
\right)\,.
\enann
Using 
$\rho_m=3\Omega_{m,0}\mpl^2 H_0^2(1+z)^3$,
we obtain
\beann
\phi=\phi_0\pm {1\over |\alpha_2|}\left[2(H(z) -H_0)-3\Omega_{m,0} H_0^2
\int_0^z dz {(1+z)^2\over H(z)}\right]
\,.
\enann
Once we know $H(z)$, we can integrate this equation, which gives the relation between $\phi$ and $z$.  Solving the inverse problem, we find the appropriate potential $V(\phi)$.

We can easily check it by assuming $\Lambda$CDM model
\beann
H^2={1\over 3M_{\rm F}^2}\left(\rho_m+V_0\right)
\,.
\enann
Since 
$\rho_m=3\Omega_{m,0}M_{\rm F}^2 H_0^2(1+z)^3$,
\beann
H^2={V_0\over 3M_{\rm F}^2}\left(1+{3\Omega_{m,0}M_{\rm F}^2 H_0^2\over V_0}(1+z)^3\right)
\,.
\enann
We then find the solution as
\beann
\phi(z)=\phi_0\pm{2\over |\alpha_2|}\left(1-{M_{\rm F}^2\over \mpl^2}\right)\left(H(z)- H_0\right)\,.
\enann
The potential is then given as
\bea
V&=&3\mpl^2H^2-\rho_m=3\mpl^2H^2-(3M_{\rm F}^2H^2-V_0) 
\nonumber 
\\
&=&3(\mpl^2-M_{\rm F}^2)H^2+V_0 
\nonumber
\\
&=&3(\mpl^2-M_{\rm F}^2)\left[H_0\pm {|\alpha_2|\over 2\left(1-{M_{\rm F}^2\over \mpl^2}\right)}(\phi-\phi_0)\right]^2+V_0
\nonumber
\\
&=&{3\alpha_2^2\mpl^4\over 4\left( \mpl^2-M_{\rm F}^2\right)}\left(\phi-\phi_*\right)^2+V_0
\label{cuscuton_LCDM1}
\,,
\ena
where
\beann
\phi_*\equiv \phi_0\mp {2H_0\left(1-{M_{\rm F}^2\over \mpl^2} \right)\over |\alpha_2|}
\,.
\enann
This is just a quadratic potential of $\phi$ with
\beann
m^2\equiv {3\alpha_2^2\mpl^4\over 2\left( \mpl^2-M_{\rm F}^2\right)}
\,,
\enann
which is consistent with Eq. (\ref{relation_MFMPL}).

\section{Exponential Potential with $\lambda\leq 6$}
\label{exponential_potential}

In \S.\ref{summary_exppot}, we give only the summary of  the cosmic evolution 
for the exponential potential
(\ref{exp_pot})
with $\lambda\leq 6$.
 In this appendix, we shall give the details of calculation. 
 The cosmic evolution can be easily understood by analyzing the behaviours of the functions 
 $D\,, S_\pm\,, R_\pm$ and $F_\pm$
 in the effective Friedmann equation
 (\ref{effective_Friedmann}).

\subsection{$0<\lambda <3$}
In this case, 
we find $S_+=0$ at $a=a_{\rm cr}$ for $+$ branch, while 
$S_-<0$ for $-$ branch.
As a result we find the following cosmic evolution:
For $+$ branch, since 
$S_+<0$  for  $a<a_{\rm cr}$ 
while $S_+>0$ for  $a>a_{\rm cr}$,
 we find
for the negative potential ($\epsilon_V=-1$), 
\beann
a_{-+}(t)  &\propto&
\left\{
\begin{array}{cccccl}
t^{1\over 2} &\rightarrow& t^{2\over 3} 
&~~~{\rm in} &{\rm the~early ~stage}
\\
 ({\rm RD})&& ({\rm 
MD})
&&
\\
&a_{\rm cr}&&~~~{\rm as}&t\rightarrow \infty
\\
\end{array}
\right. \,,
\enann
and
for the positive potential ($\epsilon_V=1$), 
\beann
a_{++}(t)  &\propto&
\left\{
\begin{array}{ccl}
a_{\rm cr}
&{\rm as} &t\rightarrow -\infty
\\
\exp[H_\infty t]
&{\rm as}&t\rightarrow \infty
\\
\end{array}
\right.
\,.
\enann

For $-$ branch, the potential must be negative ($\epsilon_V=-1$).
We then  find
\beann
 a_{--}(t)  &\propto&
\left\{
\begin{array}{cccccl}
t^{1\over 2} &\rightarrow& t^{2\over 3} 
&~~~{\rm in} &{\rm the~early ~stage}
\\
 ({\rm RD})&& ({\rm 
MD})
&&
\\
&t^{1\over 3}
&&~~~{\rm as}&t\rightarrow \infty
\\
\end{array}
\right.
\,.
\enann

Here we have used the notation for the scale factor such that $a_{\epsilon_V, {\rm branch}}$.

\subsection{$\lambda<0$}

In this case,  for $+$ branch, 
we find two vanishing points such that 
$F_+=0$ at $a=a_{\rm cr}^{(F)}$   and $S_+=0$ at $a=a_{\rm cr}^{(S)}$,
where $a_{\rm cr}^{(F)}>a_{\rm cr}^{(S)}$.
When $F_+$ vanishes, we find the Friedmann equation 
near $a_{\rm cr}^{(F)}$
as
\beann
 H^2\propto \left(a-a_{\rm cr}^{(F)}\right)^{-2}
\,,
\enann
which gives
\beann
a(t)-a_{\rm cr}^{(F)} \propto \left(t-t_{\rm cr}^{(F)} \right)^{1/2}
\,,
\enann
where $t_{\rm cr}^{(F)} $ is a positive constant.
We find a singularity at $t_{\rm cr}^{(F)} $ although the scale factor 
$a_{\rm cr}^{(F)} $ is finite.

As a result we find 
three  histories of the universe ($a_{-+}(t)\,, a_{++}^{(1)}(t) $\,,  and $a_{++}^{(2)}(t)$) as
\beann
a_{-+}(t)  &\propto&
\left\{
\begin{array}{cccccl}
t^{1\over 2} &({\rm or}~ t^{1\over 2} \rightarrow t^{2\over 3}) &
&~~~{\rm as} &t \rightarrow 0
\\
({\rm RD}) &({\rm or~RD}\rightarrow {\rm MD})&
&&
\\
a_{\rm cr}^{(S)} &&&~~~{\rm as}&t\rightarrow \infty
\\
\end{array}
\right. \,,
\\
a_{++}^{(1)}(t)  &\propto&
\left\{
\begin{array}{cccccl}
& a_{\rm cr}^{(S)}  &  
&~~~{\rm as} &t\rightarrow  -\infty
\\
&a_{\rm cr}^{(F)}&&~~~{\rm as}&t\rightarrow t_{\rm cr}^{(F)} 
\\
\end{array}
\right. \,,
\enann
and
\beann
a_{++}^{(2)}(t)  &\propto&
\left\{
\begin{array}{ccl}
a_{\rm cr}^{(F)}
&{\rm as} &t\rightarrow t_{\rm cr}^{(F)} 
\\
\exp[H_\infty t]&{\rm as}&t\rightarrow \infty
\\
\end{array}
\right.
\,.
\enann
 
For $-$ branch,
no terms vanish nor become negative, and $S_<0$.
As a result,  
for the negative potential ($\epsilon_V=-1$), we find
\beann
 a_{--}(t)  &\propto&
\left\{
\begin{array}{cccccl}
t^{1\over 2} &\rightarrow& t^{2\over 3} 
&~~~{\rm in} &{\rm the~early ~stage}
\\
 ({\rm RD})&& ({\rm 
MD})
&&
\\
&t^{1\over 3}
&&~~~{\rm as}&t\rightarrow \infty
\\
\end{array}
\right.
\,.
\enann

\subsection{$3<\lambda< 4$}
From the condition of $D\geq 0$, we find the 
lower bound on the scale factor as
\beann
a\geq a_{\rm min} \,.
\enann
For $+$ branch, we also find $a_{\rm cr} (>a_{\rm min})$ from
$S_+=0$, while for $-$ branch, no additional 
vanishing point appears.

Near $a=a_{\rm min}$, we find the Friedmann equation as
\beann
 H^2\propto \left(a-a_{\rm min}\right)
\,,
\enann
which gives
\beann
a(t)-a_{\rm min}\propto (t-t_{\rm min})^2 \,.
\enann

We then find the following cosmic evolution:
For $+$ branch, we have two histories ($a_{-+}(t)$ and $a_{++}(t)$) as
\beann
 a_{-+}(t)  &\propto&
\left\{
\begin{array}{cccccl}
a_{\rm min}
&~~~{\rm as} &t\rightarrow t_{\rm min}
\\
a_{\rm cr}&
~~~{\rm as}&t\rightarrow \infty
\\
\end{array}
\right.
\,,
\\
a_{++}(t)  &\propto&
\left\{
\begin{array}{cccccl}
a_{\rm cr}&&
&~~~{\rm as} &t\rightarrow -\infty
\\
\exp[H_\infty t]&
&&~~~{\rm as}&t\rightarrow \infty
\\
\end{array}
\right.
\,,
\enann
while for $-$ branch,
\beann
 a_{--}(t)  &\propto&
\left\{
\begin{array}{cccccl}
a_{\rm min}
&~~~{\rm as} &t\rightarrow t_{\rm min}
\\
t^{1\over 3}
&
~~~{\rm as}&t\rightarrow \infty
\\
\end{array}
\right.
\,.
\enann

\subsection{$4<\lambda<6$}
In this case, we also find the 
lower bound on the scale factor from the condition of $D\geq 0$
 as
\beann
a\geq a_{\rm min} \,.
\enann
For $-$ branch, we also find $a_{\rm cr} (>a_{\rm min})$ 
if $\lambda\geq 5$ from
$S_-=0$, while the vanishing point $S_+=0$ becomes 
 larger than  $a_{\rm min}$, which means 
$S_-$ is always negative 
for $a\leq a_{\rm min}$ if $\lambda\leq {9\over 2}$. 
In the case of ${9\over 2}<\lambda<5$, the behaviour depends on the parameters.
For $+$ branch, no additional 
vanishing point appears.

We then find the following cosmic evolution:
For $+$ branch,
\beann
 a_{++}(t)  &\propto&
\left\{
\begin{array}{cccccl}
a_{\rm min}
&~~~{\rm as} &t\rightarrow t_{\rm min}
\\
\exp[H_\infty t]
&
~~~{\rm as}&t\rightarrow \infty
\\
\end{array}
\right.
\,.
\enann
While for $-$ branch, we have two histories ($a_{+-}(t)$ and $a_{--}(t)$) as
\beann
 a_{+-}(t)  &\propto&
\left\{
\begin{array}{cccccl}
a_{\rm min}
&~~~{\rm as} &t\rightarrow t_{\rm min}
\\
a_{\rm cr}&
~~~{\rm as}&t\rightarrow \infty
\\
\end{array}
\right.
\,,
\\
a_{--}(t)  &\propto&
\left\{
\begin{array}{cccccl}
a_{\rm cr}&&
&~~~{\rm as} &t\rightarrow -\infty
\\
t^{1\over 3}&
&&~~~{\rm as}&t\rightarrow \infty
\\
\end{array}
\right.
\,.
\enann

In this case, however, we have a constraint  such that 
$
2\rho_m+\rho_r\leq V_\infty
$ from $D\geq 0$.
If $V_\infty$ is the present vacuum energy, this constraint cannot explain  
the big bang universe.

\subsection{Exponential potential with $\lambda=6$}
\label{exponential_potential_lam=6}

In this case, Eq. (\ref{potential_rho}) is a linear equation for $V$. 
Since
\beann
V\equiv \epsilon_V \mpl^4 e^{6\alpha_3\mpl^{-1}\phi}
\,,
\enann
we obtain the scalar field $\phi$ as
\bea
\phi&=&{\mpl\over 6\alpha_3}\ln
\left[
{-(\rho_m+\rho_r)+{3\over a_2^2}\left(\rho_m+{2\over 3}\rho_r\right)^2
\over
\epsilon_V \mpl^4}
\right]
\,,~~~~~~
\label{phi_lam=6}
\ena
which gives
\beann
Z&=&1+{\alpha_3\over \mpl}{d\phi\over dN} 
\\
&=&1-{\alpha_3\over \mpl}\left(3\rho_m{d\phi\over d\rho_m}+4\rho_r{d\phi\over d\rho_r}\right) 
\\
&=&{(\rho_m+{2\over 3}\rho_r)
\left(3+{4\over a_2^2}\rho_r\right)
\over 6\left[(\rho_m+\rho_r)-{3\over a_2^2}\left(\rho_m+{2\over 3}\rho_r\right)^2\right]}
\,.
\enann
Here we define
\beann
a_2\equiv {\alpha_2\over \alpha_3}\mpl^2 \,.
\enann
We then find the Friedmann equation as
\bea
\mpl^2H^2&=&Z^{-2}{\left(\rho_m+\rho_r+V(\phi)\right)\over 3} 
\nonumber \\
&=&{ 4\left[(\rho_m+\rho_r)-{3\over a_2^2}\left(\rho_m+{2\over 3}\rho_r\right)^2\right]^2\over a_2^2
\left[1+{4\over 3 a_2^2}\rho_r\right]^2} \,.
\\
&&\nonumber
\ena

If  ${\alpha_2\over \alpha_3}\gsim O(1)$, 
$\rho_m\,, \rho_r\ll a_2^2$ because $\rho_m\,, \rho_r \ll \mpl^4$. 
In this case $\epsilon_V$ must be $-1$ from Eq. (\ref{phi_lam=6}), and 
we find
\bea
\mpl^2H^2={ 4(\rho_m+\rho_r)^2\over a_2^2
}
\,,
\ena
which gives
\beann
a(t) \sim 
\left\{
\begin{array}{cc}
t^{1\over 4}&{\rm RD}
\\
t^{1\over 3}&{\rm MD}
\\
\end{array}
\right. \,.
\enann

The former expansion law is obtained by the equation of state $P={5\over 3}\rho$ in GR, which is 
quite strange matter, while the latter one corresponds to the equation of state of stiff matter.

On the other hand, if  ${\alpha_2\over \alpha_3}\ll 1$ such that $\rho_m\,,\rho_r \gg a_2^2$,
we find
\beann
\mpl^2 H^2\approx {81\over 4a_2^2}{\left(\rho_m+{2\over 3}\rho_r\right)^4\over \rho_r^2}
\,,
\enann
which gives
\beann
a(t) \sim 
\left\{
\begin{array}{cc}
t^{1\over 4}&{\rm radiation~dominant}
\\
t^{1\over 2}&{\rm matter~dominant}
\\
\end{array}
\right. \,.
\enann

The exists an intermediate parameter region such that
   $\rho_m\,,\rho_r \sim a_2^2\ll \mpl^4$.
   In this case, the Hubble expansion rate $H$ 
vanishes at some scale factor $a_{\rm cr}$ where $a_{\rm cr}$ is given by
\beann
\rho_m(a_{\rm cr})+\rho_r(a_{\rm cr})={3\over a_2^2}\left(\rho_m(a_{\rm cr})+{2\over 3}\rho_r(a_{\rm cr})\right)^2
\,.
\enann
 
 In this case, the universe expands as follows:\\
 If $\epsilon_V=-1$, we find $a\geq a_{\rm cr}$, and 
 \beann
 a(t)\sim \left\{
 \begin{array}{cc}
a_{\rm cr} & t\rightarrow -\infty
\\
t^{1\over 3}&t\rightarrow \infty
\\
\end{array}
\right.
\,,
 \enann
 while when $\epsilon_V=1$, we find $a\leq a_{cr}$ and 
  \beann
 a(t)\sim \left\{
 \begin{array}{cc}
t^{1\over 4}&t\rightarrow 0
\\
a_{\rm cr} & t\rightarrow \infty
\\
\end{array}
\right.
\,.
 \enann

\section{A negative vacuum energy}
\label{expo_potential_cosconstant}
As one of matter fluid in Eq. (\ref{matter_fluid}),
we may add a vacuum energy $\rho_{\rm vac}$.
Here we shall discuss such a case.

\begin{widetext}
 The effective Friedmann equation, when $\lambda\neq 6$,  is now :
\bea
 H^2={1\over 3\mpl^2}{V_\infty D(\tilde \rho_m, \tilde \rho_r, \tilde \rho_{\rm vac}; \lambda) S_{\pm}^2 (\tilde \rho_m, \tilde \rho_r, \tilde \rho_{\rm vac}; \lambda)
R_{\pm} (\tilde \rho_m, \tilde \rho_r, \tilde \rho_{\rm vac}; \lambda) \over 
2 F_{\pm}^2(\tilde \rho_m, \tilde \rho_r, \tilde \rho_{\rm vac}; \lambda) }
\,,
\label{effective_Friedmann_V0}
\ena
where
\bea
D(\tilde \rho_m, \tilde \rho_r, \tilde \rho_{\rm vac}; \lambda)&\equiv &
1+{4\over \lambda-6}
\left[(\lambda-3)\tilde \rho_m+(\lambda-4)\tilde \rho_r+\lambda\tilde \rho_{\rm vac}\right] \,,
\\
 S_{\pm}(\tilde \rho_m, \tilde \rho_r, \tilde \rho_{\rm vac}; \lambda)&\equiv &1 +{2\over \lambda-6}\left(3\tilde \rho_m+2\tilde \rho_r+6\tilde \rho_{\rm vac}\right)
\pm \sqrt{D} \,,
 \\
R_{\pm}(\tilde \rho_m, \tilde \rho_r, \tilde \rho_{\rm vac}; \lambda)&\equiv & 1 +{2\over \lambda-6}\left[(\lambda-3)\tilde \rho_m+(\lambda-4)\tilde \rho_r+\lambda \tilde \rho_{\rm vac}\right]
\pm \sqrt{D} \,,
\\
F_{\pm}(\tilde \rho_m, \tilde \rho_r, \tilde \rho_{\rm vac}; \lambda)&\equiv &\left\{1+{2\over \lambda(\lambda-6)}\left[3(\lambda-3)\tilde \rho_m+2(\lambda-4)\tilde \rho_r+6\lambda \tilde \rho_{\rm vac}\right]\right\}\sqrt{D}
\nn
&&
\pm\left\{1
+{2\over \lambda(\lambda-6)}\left[(\lambda-3)(2\lambda-3)\tilde \rho_m+2(\lambda-2)(\lambda-4)\tilde \rho_r+2\lambda^2\tilde \rho_{\rm vac}\right]\right\}
\,.
\ena
\end{widetext}
Here we define 
\beann
V_\infty\equiv {3\alpha_2^2\over (\lambda-6)^2\alpha_3^2}\mpl^4
\,,
\enann
and introduce the variables normalized by $V_\infty$ as
\beann
\tilde \rho_m={\rho_m\over V_\infty}\,,~
\tilde \rho_r={\rho_r\over V_\infty}\,,~
\tilde \rho_{\rm vac}={\rho_{\rm vac}\over V_\infty}
\,.
\enann

In order to find an accelerating universe in the limit of $\rho_m, \rho_r \rightarrow 0$, we find
\beann 
\tilde \rho_{\rm vac}&>-{\lambda-6\over 4\lambda}&~~~{\rm for}~\lambda>6~{\rm or}~\lambda<0 
\\
&<-{\lambda-6\over 4\lambda}&~~~{\rm for}~0<\lambda<6
\,.
\enann

The observed dark energy density is given by
\beann
\rho_{\rm DE}&\equiv &3\mpl^2 H_\infty^2 
\\
&=&
{V_\infty\over 2}
\left[
1+{2\lambda\over \lambda-6}\tilde \rho_{\rm vac}
+\sqrt{1+{4\lambda\over \lambda-6}\tilde \rho_{\rm vac}}
\right]
\,,
\enann
where 
\beann
 H_\infty\equiv H(a\rightarrow \infty)
 \,.
\enann

As  discussed in the text, $\lambda>6$ may provide a consistent cosmological history, that is,
starting from radiation era, the universe evolves into matter dominant stage, and eventually 
transits to dark energy dominant phase.
In that case, we find 
\beann
{1\over 4} V_\infty \leq\rho_{\rm DE}\leq V_\infty \,,
\enann
for 
\beann
-{\lambda-6\over 4\lambda}V_\infty \leq \rho_{\rm vac} \leq 0
\,.
\enann

A negative vacuum energy reduces dark energy density maximally to one quarter of 
the case without a negative vacuum energy.
Such a small negative cosmological constant might be obtained in the context of string theory \cite{Demirtas:2021ote}.

\section{Peculiarity of vacuum case}
If we consider there exist no matter fluid, we find some peculiarity.
In the case of the vacuum state, we have the constraint
 \bea
{1\over 3}V(\phi)=  {\alpha_3^2 \over \alpha_2^2\mpl^4 }\left[2V(\phi)- {\mpl \over 3\alpha_3}V_{,\phi}\right]^2 \,.
\label{constraint_vac}
\ena
Once we specify the potential form, this constraint fixes the value of the scalar field 
$\phi=\phi_{\rm vac} = $ constant.
Since the scalar field must be time-dependent such that  $X>0$, 
such a solution is not allowed. 
There is no vacuum solution in the Cuscuton gravity theory.
\footnote{ It is not the case if the 3-space has a curvature. 
In fact, we find de Sitter solution or Minkowski spacetime for 
the open or closed FLRW metric ansatz. $\phi$ becomes time-dependent.}

However there is one exceptional case, i.e., if the potential $V$ satisfies 
the constraint (\ref{constraint_vac}) for any value of $\phi$, it does not fix the value of $\phi$. 
Instead we find a very peculiar behaviour of the cosmic evolution or dynamics of the scalar field as shown below.
\subsection{Ordinary Cuscuton theory ($\alpha_3=0$)}
In this case, the constraint (\ref{constraint_vac}) is now
\beann
V= {1\over 3 \alpha_2^2\mpl^2 }\left(V_{,\phi}\right)^2
\label{constraint_vac1}
\,,
\enann
which gives
\beann
{dV\over d\phi}=\pm \sqrt{3}|\alpha_2|\mpl V^{1/2}
\,.
\enann
Solving this differential equation, we find the potential form as
\bea
V={3\over 4}\alpha_2^2\mpl^2(\phi-\phi_0)^2
\label{cuscuton_vac_potential}
\,.
\ena
This looks very similar to the potential for $\Lambda$CDM model given
by Eq. (\ref{cuscuton_LCDM1}). 
But in this case, $M_{\rm F}=0$ and $V_0=0$.

The evolution of the scalar field is given by
\bea
\phi=\phi_0\pm {2\over |\alpha_2|}\left(H-H_0\right)
\,,
\label{cuscuton_phi_evolution}
\ena
and the Friedman  equation is 
\bea
H^2={1\over 3\mpl^2}V(\phi)\,.
\label{cuscuton_H_evolution}
\,
\ena
Since these two equations are not independent when the potential is given by
Eq.  (\ref{cuscuton_vac_potential}), 
we cannot fix the scalar field $\phi$ or the Hubble parameter $H$.
When $H$ is given by some function of the $e$-folding number $N$, the scalar field evolves as
Eq. (\ref{cuscuton_phi_evolution}), while if we assume the evolution of $\phi$,
we find the cosmic evolution $H$ by Eq. (\ref{cuscuton_H_evolution}).
The theory cannot determine the evolution of the universe. 

What is the origin of this ambiguity or freedom ?
It may be related to a choice of the time slicing.
When we have matter fluid in the FLRW spacetime, we have a natural 
choice of time coordinate, by which the energy density becomes homogeneous.
However, if we do not have such a reference object, we may have a freedom to choose 
time coordinate, which corresponds to the above ambiguity.

\subsection{Cuscuta-Galileon  theory ($\alpha_3\neq 0$)}
We also find the similar problem for the Cuscuta-Galileon theory.
If the constraint (\ref{constraint_vac}) is  satisfied for any value of $\phi$, 
it gives the differential equation for $V(\phi)$ 
in terms of $\phi$, i.e.,
\bea
{dV\over d\phi}=6\alpha_3 \mpl^{-1} V\pm \sqrt{3}\alpha_2 \mpl V^{1/2}
\,.
\label{cuscuta_galileon_vac}
\ena
This can be easily integrated as
\beann
V(\phi)=V_0\left[1-{C\over \sqrt{3}\alpha_2}\exp\left({3\alpha_3\over \mpl}\phi\right)\right]^2
\,,
\enann
where 
\bea
V_0\equiv {\alpha_2^2\mpl^4\over 12\alpha_3^2} \,,
\label{V0}
\ena
and  $C$ is a positive  integration constant.
We shall  rewrite the potential as
\bea
V= V_0\left[1- \exp\left({3\alpha_3\over \mpl}(\phi-\phi_0)\right)\right]^2 \,.
\label{potential1}
\ena
This  is quite similar to the potential appeared 
 in the Starobinsky inflation model \cite{Starobinsky:1980te}
or the Higgs inflation model \cite{Bezrukov:2007ep,Bezrukov:2008ej,Bezrukov:2009db,DeSimone:2008ei} after conformal transformation \cite{Maeda:1987xf,Futamase:1987ua,Maeda:1988rb},
although the present scalar field is not dynamical.
The potential  approaches a positive constant as $\phi\rightarrow -\infty$,
 and vanishes at $\phi=\phi_0$, and then it increases and 
 diverges as
$\phi\rightarrow \infty$.

In this case, we also find one independent equation for two unknown variables $\phi$ and $H$,
which is 
\beann
\epsilon_{\dot \phi}H
={\alpha_2\mpl\over 6\alpha_3}{\left(1-\exp\left[{3\alpha_3\over \mpl}(\phi -\phi_0)
\right]\right)\over \left(1+{\alpha_3\over \mpl}{d\phi\over dN}\right)}
\,.
\enann
For given arbitrary function of $\phi(N)$, we find the evolution of the universe given by this Hubble parameter $H$, or vice versa.

\subsection{Case with matter field}
In the case of the original Cuscuton gravity, if the potential $V$ is given by
Eq. (\ref{cuscuton_vac_potential}), 
we cannot introduce matter fluid.
The basic equations force matter density to $0$.

On the other hand, for the Cuscuta-Galileon gravity, 
the situation changes.
We can add matter fluid in the Cuscuta-Galileon theory with the potential (\ref{potential1}).
We shall discuss its cosmic evolution. 

If we assume the potential $V$ is given by Eq. (\ref{potential1}),
the constraint (\ref{constraint}) becomes
\beann
{\rho\over 3}=  {\left(\rho-P \right)\over a_2^2}\left[\left(\rho-P \right)+2\left(2V-{\mpl\over 3\alpha_3}V_{,\phi}\right)
\right] \,,
\label{constraint2}
\enann
In this case, there are two branches: One is vacuum ($\rho=P=0$),
and the other gives
\bea
V=V_0{\left[\rho-{1\over 4V_0}\left(\rho-P\right)^2
\right]^2\over \left(\rho-P\right)^2}
\,.
\ena
Here we use the condition (\ref{cuscuta_galileon_vac}) and the definition (\ref{V0}), i.e., 
\beann
2V-{\mpl\over 3\alpha_3}V_{,\phi}=\pm 2\sqrt{V_0V} \,,
\enann
to eliminate $V_{,\phi}$.

Assuming $\rho=\rho_m$, we find
\beann
{V\over V_0}=\left(1-{\rho_m\over 4V_0}\right)^2
\,.
\enann
Since the potential $V$ is given by the scalar field $\phi$ as Eq. (\ref{potential1}),
this equation 
 determines the behaviour of $\phi$ in terms of $\rho_m$
\,, i.e.,
\bea
\exp\left[{3\alpha_3\over \mpl}
 (\phi-\phi_0)\right]&=&1\mp \left(1-{\rho_m\over 4V_0}\right)
\nonumber
\\
&=&
\left\{
\begin{array}{c}
\displaystyle{{\rho_m\over 4V_0}}
\label{case1}
\\[1em]
\displaystyle{2-{\rho_m\over 4V_0}}
\label{case2}
\end{array}
\right. \,.
\ena
We find two solution for $\phi$ as
\bea
\phi=\phi_\mp\equiv \left\{
\begin{array}{l}
\phi_0+\displaystyle{{\mpl\over 3\alpha_3}\ln {\rho_m\over 4V_0}}
\label{case1}
\\[1em]
\phi_0+\displaystyle{{\mpl\over 3\alpha_3}\ln {\left(2-{\rho_m\over 4V_0}\right)}}
\label{case2}
\end{array}
\right. \,.
\ena
The Friedmann equation is now
\bea
\bar H^2=H^2 Z^2 &=&{1\over 3\mpl^2}\left(\rho_m+V\right)\nn
&=&{1\over 3\mpl^2}\left[\rho_m+V_0\left(1-{\rho_m\over 4V_0}\right)^2\right]
\nn
&=&{V_0\over 3\mpl^2}\left(1+{\rho_m\over 4V_0}\right)^2
\,,
\label{1stFriedmanneq}
\ena
where 
\beann
Z=1+{\alpha_3\over \mpl}{d\phi\over dN}
\,.
\enann
Since $\rho_m\propto e^{-3N}$, we find
\beann
Z(\phi_-)=1+{\alpha_3\over \mpl}{d\over dN}\left({\mpl\over 3\alpha_3}\ln {\rho_m\over 4V_0}\right)=0\,,
\enann
which is an irrelevant solution. 
For $\phi_+$, we find
\beann
Z(\phi_+)&=&1+{\alpha_3\over \mpl}{d\over dN}\left({\mpl\over 3\alpha_3}\ln \left(2-{\rho_m\over 4V_0}\right)\right)
\\
&=&
{1\over \displaystyle{1-{\rho_m\over 8V_0}}}\,.
\enann

We obtain  the Friedmann equation (\ref{1stFriedmanneq}) as
\beann
H
&=&H_{\rm vac}\left(1-{\rho_m\over 8V_0}\right)\left(1+{\rho_m\over 4V_0}\right)
\,,
\enann
where
\bea
H_{\rm vac}\equiv \sqrt{V_0\over 3} \frac{1}{\mpl} = {|\alpha_2|\over 6\alpha_3}\mpl
\,.
\ena
Note that $\rho_m\leq 8V_0$, which strongly restricts matter density.

Introducing 
\beann
\eta\equiv {\rho_m\over 8V_0}
\,,
\enann
which is proportional to $e^{-3N}$,
we find 
\beann
H={dN\over dt}=-{1\over 3}{d\ln \eta\over dt}\,.
\enann
The Friedmann equation is now
\beann
-{1\over 3}{d\ln \eta\over dt}
=H_{\rm vac}\left(1-\eta\right)\left(1+2\eta \right)
\,.
\enann
We can easily integrate this equation as
\beann
\ln {\eta\over (1-\eta)^{1/3}(1+2\eta)^{2/3}}=-3H_{\rm vac}(t-t_*)\,,
\enann
or 
\bea
{\eta\over (1-\eta)^{1/3}(1+2\eta)^{2/3}}
=\exp\left[-3H_{\rm vac}(t-t_*)\right]
 \label{sol_eta}
 \,,
\ena
where $t_*$ is an integration constant.
This solution gives the time evolution of matter density as
\beann
\rho_m=8V_0 \eta(t)
\,.
\enann
and the behaviour of the scale factor as
\beann
a=a_0\left({\eta\over \eta_0}\right)^{-{1\over 3}}
\,.
\enann

In order to find the explicit form, we have to solve the cubic
equation (\ref{sol_eta}) for $\eta$.

We consider some limiting cases as follows:
\\[1em]
(1)~$\eta \rightarrow 0$ \\[.5em]
This limit corresponds to $\rho_m\rightarrow 0$ or  $a\rightarrow \infty$.
We find from Eq. (\ref{sol_eta})
\beann
{1\over \eta }\propto \exp\left[3H_{\rm vac} t\right]\,,
\enann
and 
\beann
a\propto \exp\left[H_{\rm vac}  t\right] \,.
\enann
The scalar field approaches some constant as
\beann
\phi\rightarrow \phi_0 +{\mpl\over 3\alpha_3}
\ln 2
\,.
\enann
The potential value approaches 
as
\beann
V\rightarrow V_0
\,.
\enann
We find de Sitter accelerating universe.\\[1em]
(2)~$\eta \rightarrow 1$\\[.5em]
In this limit, which corresponds to  $\rho_m\rightarrow 8V_0$ and $a\rightarrow$ constant, we find 
\beann
\eta \rightarrow 1-{1\over 9}\exp\left[9H_{\rm vac} (t-t_*)\right]
\enann
as 
\beann
t\rightarrow -\infty\,.
\enann
The scalar field behaves as
\beann
\phi\rightarrow -\infty
\,.
\enann

\noindent
(3) Whole history \\[.5em]
We then find the evolution of the universe as follows: 
\bea
a(t)\propto 
\left\{
\begin{array}{ccc}
{\rm constant}&{\rm as}&t\rightarrow -\infty 
\\
\exp (H_{\rm vac}t)&{\rm as}&t\rightarrow \infty
\\
\end{array}
\right. \,,
\ena
\bea
\phi\propto 
\left\{
\begin{array}{ccc}
-\infty &{\rm as}&t\rightarrow -\infty 
\\
 \phi_0 +{\mpl\over 3\alpha_3}\ln 2&{\rm as}&t\rightarrow \infty
\\
\end{array}
\right. \,,
\ena
\bea
V=
\left\{
\begin{array}{ccc}
V_0&{\rm as}&t\rightarrow -\infty 
\\
V_0&{\rm as}&t\rightarrow \infty
\\
\end{array}
\right. \,.
\ena

There is no matter/radiation dominant stage.
This can be easily understood from the fact that
\beann
\rho_m\leq 8V_0
\,.
\enann

\hskip 1cm


%

\end{document}